\documentclass[aps,prl,twocolumn,superscriptaddress]{revtex4-2}

\usepackage{dcolumn}
\usepackage{siunitx}
\usepackage{graphicx}

\newcolumntype{d}[1]{D{.}{.}{#1}}

\begin{document}


\title{Quantum Electrodynamics in Strong Electromagnetic Fields:\\Substate Resolved K$\alpha$ Transition Energies in Helium-like Uranium}

\newcommand{\HIJena}{Helmholtz Institute Jena, 07743 Jena, Germany}
\newcommand{\GSIDarmstadt}{GSI Helmholtzzentrum für Schwerionenforschung GmbH, 64291 Darmstadt, Germany}
\newcommand{\IOQJena}{Institute of Optics and Quantum Electronics,\\ Friedrich Schiller University Jena, 07743 Jena, Germany}
\newcommand{\KIPHeidelberg}{Kirchhoff Institute for Physics, Heidelberg University, 69120 Heidelberg, Germany}
\newcommand{\JLUGiessen}{I. Physikalisches Institut, Justus-Liebig-Universität Gießen, 35392 Gießen, Germany}
\newcommand{\LKBParis}{Laboratoire Kastler Brossel, Sorbonne Université, 75005 Paris, France}
\newcommand{\HFHFGiessen}{Helmholtz Forschungsakademie Hessen für FAIR (HFHF), Campus Gießen, 35392 Gießen, Germany}
\newcommand{\SPAEdinburgh}{School of  Physics and Astronomy, The University of Edinburgh, EH9 3FD  Edinburgh, UK}
\newcommand{\SMEEChuzhou}{School of Mechanical and Electrical Engineering, Chuzhou University, 239000 Chuzhou, China}
\newcommand{\MSIKrakow}{Marian Smoluchowski Institute of Physics, Jagiellonian University, 30-348 Kraków, Poland}


\author{Ph.~Pfäfflein}
\email[]{p.pfaefflein@hi-jena.gsi.de}
\affiliation{\HIJena}
\affiliation{\GSIDarmstadt}
\affiliation{\IOQJena}

\author{G.~Weber}
\affiliation{\HIJena}
\affiliation{\GSIDarmstadt}

\author{S.~Allgeier}
\affiliation{\KIPHeidelberg}

\author{Z.~Andelkovic}
\affiliation{\GSIDarmstadt}

\author{S.~Bernitt}
\affiliation{\HIJena}
\affiliation{\GSIDarmstadt}

\author{A.~I.~Bondarev}
\affiliation{\HIJena}
\affiliation{\GSIDarmstadt}

\author{A.~Borovik, Jr.}
\affiliation{\JLUGiessen}
\affiliation{\HFHFGiessen}

\author{L.~Duval}
\affiliation{\LKBParis}

\author{A.~Fleischmann}
\affiliation{\KIPHeidelberg}

\author{O.~Forstner}
\affiliation{\HIJena}
\affiliation{\GSIDarmstadt}
\affiliation{\IOQJena}

\author{M.~Friedrich}
\affiliation{\KIPHeidelberg}

\author{J.~Glorius}
\affiliation{\GSIDarmstadt}

\author{A.~Gumberidze}
\affiliation{\GSIDarmstadt}

\author{Ch.~Hahn}
\affiliation{\HIJena}
\affiliation{\GSIDarmstadt}

\author{F.~Herfurth}
\affiliation{\GSIDarmstadt}

\author{D.~Hengstler}
\affiliation{\KIPHeidelberg}

\author{M.~O.~Herdrich}
\affiliation{\HIJena}
\affiliation{\GSIDarmstadt}
\affiliation{\IOQJena}

\author{P.-M.~Hillenbrand}
\affiliation{\GSIDarmstadt}
\affiliation{\JLUGiessen}
\affiliation{\HFHFGiessen}

\author{A.~Kalinin}
\affiliation{\GSIDarmstadt}

\author{M.~Kiffer}
\affiliation{\HIJena}
\affiliation{\GSIDarmstadt}
\affiliation{\IOQJena}

\author{F.~M.~Kröger}
\affiliation{\HIJena}
\affiliation{\GSIDarmstadt}
\affiliation{\IOQJena}

\author{M.~Kubullek}
\affiliation{\IOQJena}

\author{P.~Kuntz}
\affiliation{\KIPHeidelberg}

\author{M.~Lestinsky}
\affiliation{\GSIDarmstadt}

\author{Yu.~A.~Litvinov}
\affiliation{\GSIDarmstadt}

\author{B.~Löher}
\affiliation{\GSIDarmstadt}

\author{E.~B.~Menz}
\affiliation{\HIJena}
\affiliation{\GSIDarmstadt}
\affiliation{\IOQJena}

\author{T.~Over}
\affiliation{\HIJena}
\affiliation{\IOQJena}

\author{N.~Petridis}
\affiliation{\GSIDarmstadt}

\author{S.~Ringleb}
\affiliation{\HIJena}
\affiliation{\IOQJena}

\author{R.~S.~Sidhu}
\affiliation{\GSIDarmstadt}
\affiliation{\SPAEdinburgh}

\author{U.~Spillmann}
\affiliation{\GSIDarmstadt}

\author{S.~Trotsenko}
\affiliation{\GSIDarmstadt}

\author{A.~Warczak}
\affiliation{\MSIKrakow}

\author{B.~Zhu}
\affiliation{\SMEEChuzhou}

\author{Ch.~Enss}
\affiliation{\KIPHeidelberg}

\author{Th.~Stöhlker}
\affiliation{\HIJena}
\affiliation{\GSIDarmstadt}
\affiliation{\IOQJena}

\date{\today}

\begin{abstract}
Using novel metallic magnetic calorimeter detectors at the CRYRING@ESR, we recorded X-ray spectra of stored and electron cooled helium-like uranium (U$^{90+}$) with an unmatched spectral resolution of close to $90$\,eV. This allowed for an accurate determination of the energies of all four components of the K$\alpha$ transitions in U$^{90+}$. We find good agreement with state-of-the-art bound-state QED calculations for the strong-field regime. Our results do not support any systematic deviation between experiment and theory in helium-like systems, the presence of which was subject of intense debates in recent years.
\end{abstract}


\maketitle


The study of electrons in extreme electromagnetic fields, as present in heavy highly-charged ions (HCI), is one of the frontiers in exploring quantum electrodynamics (QED). At high atomic numbers $Z$, the electron–nucleus coupling constant $\alpha Z$ approaches unity, where $\alpha \approx 1/137$ is the fine structure constant. Therefore, for heavy HCI, perturbative treatments in $\alpha Z$, that enable highly accurate QED predictions for hydrogen and other light systems, are no longer applicable~\cite{Shabaev2018}. Nowadays, challenging non-perturbative calculations in $\alpha Z$ can be performed up to the second order of the expansion in $\alpha$, including one-electron two-loop contributions as well as two-electron QED effects~\cite{Yerokhin2015, Artemyev2005, Malyshev2019, Kozhedub2019}.

The most stringent experimental tests of bound-state QED in the presence of strong electromagnetic fields are provided by measurements of transition energies~\cite{Loetzsch2024}, hyperfine structure~\cite{ullmann2017, skripnikov2018} and bound electron g-factors~\cite{Morgner2023}. For the heaviest HCI, up until now, measurements of transition energies have delivered the most sensitive tests of various QED contributions. For hydrogen-like and lithium-like systems, several high-precision measurements have already been performed, as summarized by Beiersdorfer~\cite{Beiersdorfer2010} and Indelicato~\cite{Indelicato2019}. In contrast, precision spectroscopy of helium-like ions heavier than xenon ($Z=54$) have been virtually non-existent until very recently. It is important to emphasize that helium-like ions represent the simplest multi-electron systems where a complex interplay between relativistic, electron--electron correlation and QED effects can be explored. QED calculations of the binding energies in helium-like ions for principal quantum numbers $n\,=\,1$ and $n\,=\,2$ have been accomplished for $Z=12\,\text{--}\,100$ by Artemyev~et~al.~\cite{Artemyev2005}. Refinements of these calculations have recently been reported for specific nuclei by Malyshev~et~al.~\cite{Malyshev2019} and Kozhedub~et~al.~\cite{Kozhedub2019}. Furthermore, a successful validation of QED predictions in this regime of very strong coupling may provide novel opportunities for tests of fundamental physics and searches for physics beyond the Standard Model~\cite{kozlov_highly_2018,safronova_search_2018,Safronova2019}.  Recently, groundbreaking results on the intra-shell transition $1 \text{s} 2 \text{p} \text{ }^3 \text{P}_2 \text{ } \rightarrow \text{ } 1 \text{s} 2 \text{s} \text{ }^3 \text{S}_1$ (close to 4.5\,keV) were reported by Loetzsch~et~al.~\cite{Loetzsch2024}, confirming theoretical predictions of the screened and two-loop one-electron QED contributions in helium-like heavy ions for the first time.

\begin{figure}[tbh]
\includegraphics[]{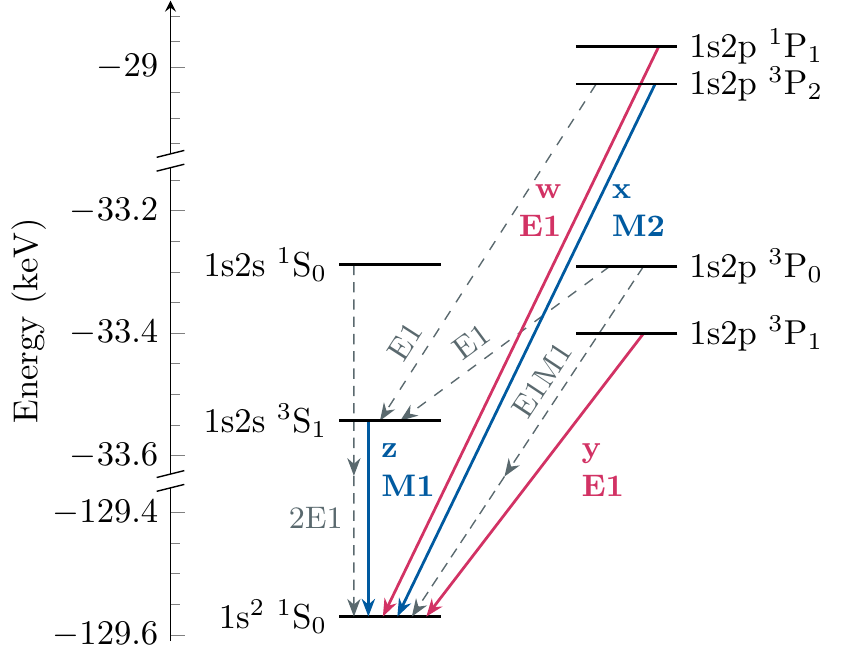}
\caption{\label{fig:scheme}Level scheme of helium-like uranium, where the most prominent radiative transitions are depicted. The K$\alpha$ transitions studied in this work are drawn in magenta and blue.}
\end{figure}

However, a similar test is still missing for the K$\alpha$ transitions ($n\,=\,2 \rightarrow n\,=\,1$) which provide access to the most strongly bound states being most affected by QED effects. In the heaviest systems, precision spectroscopy of this radiation is particular demanding due to the high X-ray energies close to 100\,keV in combination with the presence of closely spaced L-shell sublevels, as presented in Fig.~\ref{fig:scheme}. Disentangling the individual transitions that form the K$\alpha$ peaks, called w, x, y and z according to Gabriel's notation~\cite{Gabriel1972}, is hardly possible, due to the limited spectral resolution of commonly used semiconductor detectors. In U$^{90+}$, the closest lines w and x are separated by 74\,eV compared to a typical energy resolution of a few 100\,eV full width at half maximum (FWHM). Consequently, precision spectroscopy of K$\alpha$ transitions has so far been performed only up to xenon~\cite{Thorn2009}. The need for accurate measurements in high-$Z$ systems is particular pressing as possible $Z$-dependent deviations between experiment and theory have been intensely discussed~\cite{Chantler2012,Epp2013,Chantler2013,Payne2014,Kubicek2014,Chantler2014,Epp2015,Beiersdorfer2015,Machado2018,Indelicato2019}.

In this Letter, we present a study using novel high-resolution detectors on helium-like uranium (U$^{90+}$) at the GSI/FAIR facility in Darmstadt, Germany. So far, the most precise studies of these transitions were performed using semiconductor detectors~\cite{Briand90,Gumberidze2004}. The limited resolution provided by this detector technology hindered accurate comparison of the observed transition energies with theoretical predictions. In contrast, in the present work we used an alternative detector technology featuring a superior spectral performance, similarly to the study of helium-like xenon~\cite{Thorn2009}. At the electron cooler of the recently installed CRYRING@ESR~\cite{lestinsky_first_2022}, we employed two maXs-type detectors (Micro-Calorimeter Arrays for High Resolution X-ray Spectroscopy)~\cite{Pies2012,Hengstler2015}. A spectral resolution of $\leq 90$\,eV FWHM over a wide energy range from a few keV to above 100\,keV was achieved. This enabled to disentangle all four components of the K$\alpha$ radiation of U$^{90+}$, resulting in an accurate determination of their transition energies, without relying on assumptions about the relative line intensities. As a cross-check, we also obtained an accurate value of the U$^{90+}$ ground-state ionization energy. While a comprehensive description of the experimental setup and its operation is provided in~\cite{Pfaefflein2022}, in the following we briefly highlight key features of the experiment before proceeding to the discussion of the obtained results.


Excited U$^{90+}$ ions were produced from a primary beam of hydrogen-like U$^{91+}$ ions via radiative recombination in the electron cooler of CRYRING@ESR. Starting with a beam of U$^{91+}$ at a kinetic energy of 296\,MeV/u, in the ESR successive electron cooling and deceleration steps were applied. At an energy of 10.225\,MeV/u the ions were transferred to the low-energy storage ring CRYRING@ESR. This multi-staged preparation procedure resulted in one injection every 55\,s. The experiment was conducted with an average of about $1 \times 10^6$ U$^{91+}$ ions per injection over a week of continuous measurement time.

In CRYRING@ESR, the stored ions were continuously electron-cooled with a voltage of 5634.5\,V applied to the cooler. At an electron current of 30.5\,mA an electron density of approximately $ 1.2\,\times\,10^{7}\,\text{cm}^{-3}$ was reached. The repulsive force by the space charge of the electron beam reduced the acceleration potential to 5609.4\,V. We assume an uncertainty of $\pm \, 2$\,V to account for the unknown contact potential between the cathode and collector electrode of the electron cooler. This translates to an expected ion beam velocity of $\beta = 0.14695 (3)$. 

At relative velocities between electrons and ions close to zero, as is the case during electron cooling, the recombination process predominantly captures electrons into Rydberg states. The excited ions subsequently decay to the ground state via radiative cascades. Thus, each recombination event typically results in the emission of numerous characteristic photons in addition to the radiative recombination photon, as discussed in detail for example by Zhu et~al.~\cite{Zhu2022}. The U$^{90+}$ ions were separated from the main beam in the next dipole magnet and then registered by a particle detector.

The cooler section features view ports located at 0$^\circ$ and 180$^\circ$ with respect to the ion beam axis, that are equipped with 50\,\textmu m thin beryllium X-ray windows. At these angles the Doppler shift gets virtually independent of the observation angle. Therefore the influence of a slight detector misalignment on the recorded photon energy is negligible, see~\cite{Kroeger23} for details. This is in sharp contrast to crystal spectrometer setups with detection perpendicular to the ion beam axis~\cite{Gassner2018, Loetzsch2024}, which require a precise knowledge of the geometry. At both view ports a maXs-type microcalorimeter was placed. The resulting distance between the detectors and the center of the interaction zone of stored ions and cooler electrons was approximately 3.5\,m. 

\begin{figure}[tbh]
\includegraphics[]{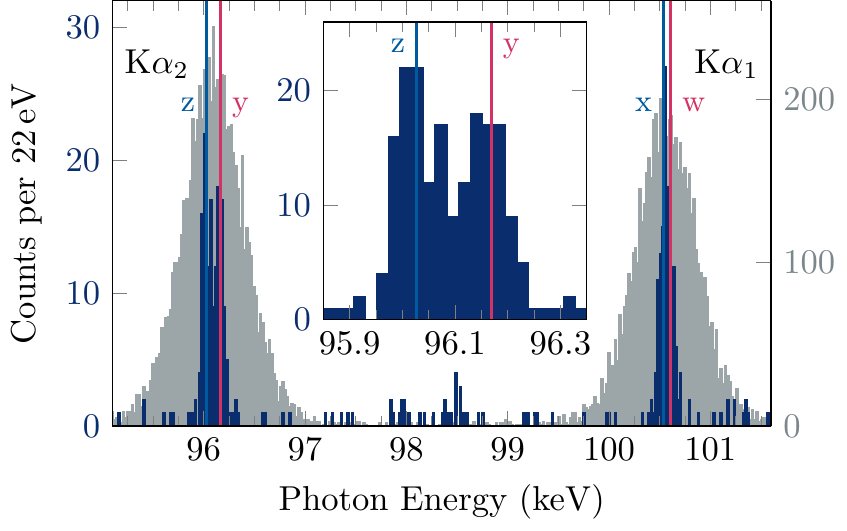}
\caption{\label{fig:fit}Spectral distribution of the K$\alpha$ transitions in U$^{90+}$, with photon energies transformed into the emitter frame. Dark blue: Data of the present study with the spectra of both microcalorimeters added up, counts on left y-axis. As shown in the inset, the lines z and y have been resolved for the first time. Light grey: Spectrum recorded with a germanium detector (data from \cite{Gumberidze2004}), counts according to right y-axis.}
\end{figure}

The detectors are based on the metallic magnetic calorimeter (MMC) technology. The energy of an incident particle is converted into thermal energy in a small absorber. The resulting temperature rise leads to a decrease of the magnetization of a thermally coupled paramagnetic sensor made of sputtered Ag:Er. The changing magnetization is in turn measured by a low-noise, high-bandwidth superconducting quantum interference device (SQUID) magnetometer. The necessary temperatures for operating these quantum sensors of below 20\,mK can be achieved using $^{3}$He/$^{4}$He dilution refrigerator cryostats. This technology allows for a unique combination of a high spectral resolution with a broad bandwidth acceptance~\cite{Kempf2018}.

In the present study, two identical prototypes of the maXs-100 design were employed. They feature an $8 \times 8$ array of absorbers made of gold. Each absorber has an area of 1.25\,mm $\times$\,1.25\,mm and a thickness of 50\,\textmu m. These detectors are tailored for a spectral resolution of $\Delta E_{\text{FWHM}}\,<\,50$\,eV in a broad photon energy range from a few keV to above 100\,keV. The design provides a photopeak efficiency of higher than 10\,\% up to 130\,keV photon energy. Furthermore, the maXs detectors' fast signal rise time was exploited for the first time in this measurement to extract timing information. Setting a coincidence condition on the simultaneous detection of photons and down-charged U$^{90+}$ ions suppressed the background by two orders of magnitude~\cite{Pfaefflein2022_coinc}. This procedure yielded almost background-free spectra of the radiation emitted in association with the recombination of cooler electrons with the stored U$^{91+}$ ions. 

The lifetime of the ion beam in CRYRING@ESR was $(7-8)$\,s, leaving the ring virtually empty after 25\,s. The remaining 30\,s period of the accelerator cycle was used for energy calibration of the detectors. The isotopes $^{241}$Am, $^{57}$Co, $^{109}$Cd, and $^{153}$Gd were selected as reference sources. They offer a set of well-known $\gamma$ lines in the energy range of interest. The intensity of the sources was chosen low enough so that a potential input power dependent shift as observed by Thorn et~al.~\cite{Thorn2009} was negligible. A second-order polynomial function was used to translate pulse amplitudes to incident photon energies, in accordance with the expected gain behaviour of MMC detectors~\cite{Bates2016}. For the 0$^\circ$ detector the lines near 26\,keV and 60\,keV of $^{241}$Am and the lines near 122\,keV and 136\,keV of $^{57}$Co were used for calibration. In case of the 180$^\circ$ detector the lines close to 60\,keV of $^{241}$Am, close to 88\,keV in $^{109}$Cd and close to 122\,keV in $^{57}$Co were chosen. The energy values of the lines were taken as reported by Helmer and van~der~Leun~\cite{Helmer2000}.

The use of reference lines that are tens of keV away from the lines of interest is likely to cause a sizable uncertainty in the determination of their positions. To quantify this effect, we selected the 97\,keV and 103\,keV lines of $^{153}$Gd that are sufficiently close to the energy region of interest in both detectors as a benchmark. The difference between the measured line positions and the literature values are $0.24 \pm 0.99$\,eV and $2.0 \pm 1.3$\,eV for the 0$^\circ$ detector and $-1.32 \pm 0.40$\,eV and $-2.51 \pm 0.50$\,eV for the 180$^\circ$ detector. From this, we inferred a conservative estimate of $\pm$2.5\,eV for the systematic uncertainty of energy determination in our region of interest. For the reference lines we measured instrumental line widths between 65\,eV and 90\,eV FWHM. This degradation in resolution compared to the design value of the detectors is partly attributed to the coupling of electromagnetic interferences into the SQUID readout electronics during the beamtime. Furthermore, mechanical vibrations can introduce an additional heat load into the cryostats and therefore deteriorate the resolution.


\begin{table}[tbh]
\caption{Energies of the K$\alpha$ transitions in U$^{90+}$ obtained in this work ($E_\text{exp}$) in comparison with theory ($E_\text{th}$)~\cite{Kozhedub2019}. The lines are referred to by their initial states as well as by Gabriel's notation~\cite{Gabriel1972} in parentheses. Furthermore, the natural line widths $\Delta E_{nat}$ of the initial states are shown. All energies are given in units of eV. An additional systematic uncertainty due to the calibration is estimated to be 2.5\,eV for all transitions.
\label{tab:results}} 
\begin{ruledtabular}
\begin{tabular}{rlcd{1}cd{2}lcd{1}cd{1}}
\multicolumn{2}{c}{Transition} &  & \multicolumn{1}{c}{$\Delta E_\text{nat}$} &  & \multicolumn{2}{c}{$E_\text{th}$} &  & \multicolumn{3}{c}{$E_\text{exp}$} \\
\hline
$1 \text{s} 2 \text{p} \text{ }^1 \text{P}_1$ & (w) & & 33.8 & & 100610.68 & (54) & & 100609 & $\pm$ & 13 \\
$1 \text{s} 2 \text{p} \text{ }^3 \text{P}_2$ & (x) & & 0.2 & & 100536.95 & (54) & & 100543.1 & $\pm$ & 7.6 \\
$1 \text{s} 2 \text{p} \text{ }^3 \text{P}_1$ & (y) & & 20.1 & & 96169.43 & (54) & & 96161.0 & $\pm$ & 6.7 \\
$1 \text{s} 2 \text{s} \text{ }^3 \text{S}_1$ & (z) & & 0.1 & & 96027.07 & (54) & & 96027.9 & $\pm$ & 5.0 \\
\end{tabular}
\end{ruledtabular}
\end{table}

\begin{figure*}[tbh]
\includegraphics[]{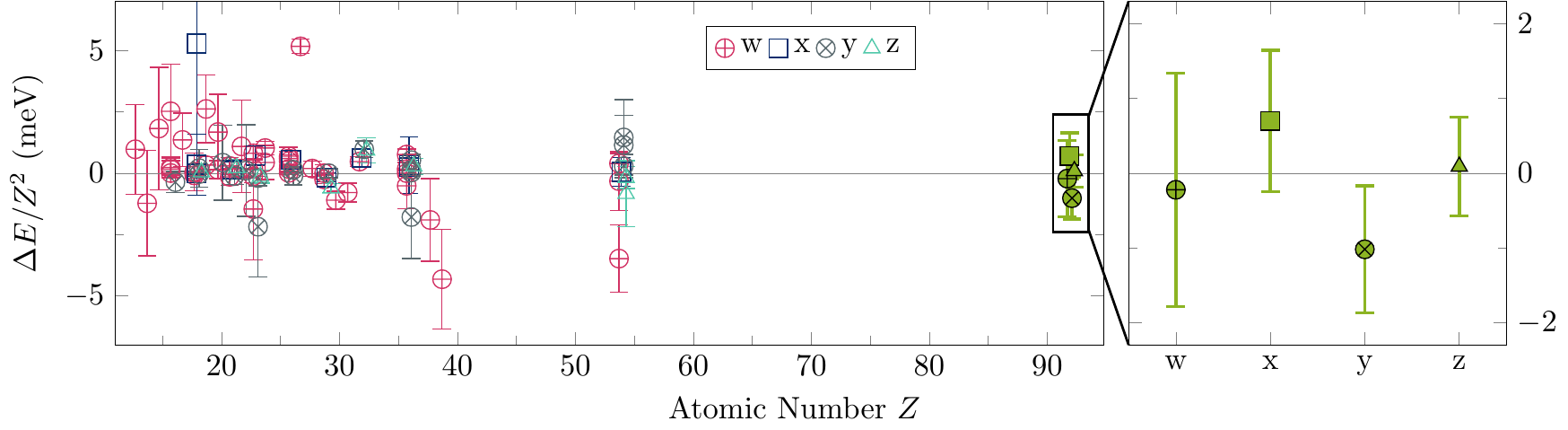}
\caption{\label{fig:deviation_or_not} Left part: Overview of available experimental data of the four K$\alpha$ transitions in helium-like systems, taken from~\cite{Indelicato2019} and references therein. Shown is the normalized difference between measured energies and theoretical predictions~\cite{Artemyev2005}. The results of the present work (shown by green symbols) provide the first accurate data for systems heavier than xenon ($Z=54$). Right part: expanded view of our data points for He-like uranium.}
\end{figure*}

In the following, we focus on the K$\alpha$ transitions in U$^{90+}$. Each detector recorded about 150 K$\alpha$ photons, whose energies were Doppler shifted to the region between 111\,keV and 117\,keV at 0$^\circ$ and 82.5\,keV to 87\,keV at 180$^\circ$, respectively. The predicted energy splittings of their components, x and w in case of K$\alpha_1$ and z and y for K$\alpha_2$, are of similar size as the detector resolution. Figure~\ref{fig:fit} shows the combined spectral data in the K$\alpha$ region from both microcalorimeters, with the photon energies transformed into the emitter frame. To highlight the improvement in resolution a spectrum previously recorded with a germanium detector (data from \cite{Gumberidze2004}) is shown as well. The achieved detector resolution proved suitable for separating the contributions of the z and y lines to the K$\alpha_2$ peak. The w and x transitions are superimposed and form a single peak (K$\alpha_1$). However, a detailed investigation of the peak shape clearly shows that it is formed by two individual lines.

To obtain the transition energies, four Voigt-shaped peaks were matched to the spectra recorded by each detector, applying an unbinned maximum-likelihood adjustment using the iminuit package~\cite{James:1975dr, iminuit}. The width parameters of the peaks were fixed: The natural line widths of up to 34\,eV, being not negligible compared to the detector resolution, were taken from theory as listed in Tab.~\ref{tab:results}. The detectors' response function was assumed to be Gaussian-shaped, with width determined from the nearest $\gamma$ reference lines to values of 81\,eV for the $0^\circ$ detector and 90\,eV for the $180^\circ$ detector. The relative intensities and the centroid positions were used as free parameters for the adjustment, together with the ion beam velocity to account for the Doppler shift. The adjustment yields $\beta = 0.146987(34)$, in good agreement with the value estimated from the electron cooler settings. This also serves as a consistency check for the energy calibration of both detectors.

Our experimental values for the transition energies are given in Tab.~\ref{tab:results}, together with theory values taken from~\cite{Kozhedub2019}. Experimental uncertainties are dominated by counting statistics and given for $1\sigma$ confidence intervals. Our measurement provides values for all four components of the K$\alpha$ transitions in U$^{90+}$ for the first time. An overview of the existing experimental data for all the K$\alpha$ transition energies for atomic numbers $Z=12\text{--}92$ is given in Fig.~\ref{fig:deviation_or_not}. Plotted is the differences between theoretical and experimental values normalized to $Z^2$. While the low- to medium-$Z$ range was extensively studied, our measurement is the first one at $Z>54$ to achieve the necessary accuracy for a meaningful test of theory. We acknowledge that published data exist also for K$\alpha$ transitions in $Z=59$~\cite{thorn_high-resolution_2008} and $Z=92$~\cite{Briand90,Lupton94}. However, in these studies the determination of the line position either relied on an assumption about the population of the excited states and the resulting relative line intensities or used theory values for some K$\alpha$ transitions as an energy reference for the others. For the case of U$^{90+}$, our measurement constitutes an improvement of up to one order of magnitude of the uncertainty compared to these previous studies. Our findings are in good agreement with theory and do not suggest any systematic deviation scaling with 4$^{\text{th}}$ or higher powers of $Z$ as proposed in~\cite{Chantler2013,Chantler2014}.

As a consistency check, we also determined the ground-state ionization energy $E_{\text{ion}}$ for U$^{90+}$, by assuming that the L-shell ionization energies are accurately known from theory~\cite{Kozhedub2019}. This assumption is supported by a recent high-precision measurement of the intra-shell transition $1 \text{s} 2 \text{p} \text{ }^3 \text{P}_2 \text{ } \rightarrow \text{ } 1 \text{s} 2 \text{s} \text{ }^3 \text{S}_1$ in U$^{90+}$~\cite{Loetzsch2024}, showing an agreement with theory on the level of $0.17$\,eV. With the binding energies of the L shell states being fixed, an adjustment was performed with the ground-state ionization energy as a free parameter $E_{t} = E_{\text{ion}} - E_{\text{L}i}$ (here $E_{t}$ with $t \in \text{x, y, w, z}$ is the transition energy and $E_{\text{L}i}$ the ionization energy of the corresponding initial L substate, see Fig.~\ref{fig:scheme}). This evaluation yields $E_{\text{ion}} = 129570.0 \pm 4.2 \pm 2.5$\,eV, which is in good agreement with the theoretical value of $129570.09(53)$\,eV taken from~\cite{Kozhedub2019}. Previously, the only experimental determination of this quantity was a relative measurement to the ground-state binding-energy of U$^{91+}$~\cite{Gumberidze2004}. Combining the relative value of $2248 \pm 9$\,eV reported there with the measured ground-state ionization energy of the hydrogen-like system of $131819.8 \pm 4.6$\,eV~\cite{Gumberidze2005} results in a value of $E_{\text{ion}} = 129571 \pm 10$\,eV, which is also in agreement with our value. 


In summary, we performed a spectral study of X-ray transitions in helium-like uranium using novel MMC detectors at the electron cooler of the CRYRING@ESR. Achieving a resolution of close to 90\,eV FWHM for photon energies of up to 150 keV is a major breakthrough for precision spectroscopy of heavy ions at storage rings. It allowed us to disentangle the components of the K$\alpha$ lines in U$^{90+}$ for the first time. The measured transition energies are in good agreement with state-of-the-art bound-state QED calculations. This finding does not support any systematic deviation that scales with 4$^{\text{th}}$ or higher powers of $Z$, as was recently proposed and intensely debated within the community.

We like to highlight that when combining MMC detectors with the beneficial experimental conditions at the CRYRING@ESR electron cooler, the only significant systematic uncertainty is the quality of the energy calibration. By optimizing the calibration procedure and using a broader set of gamma reference lines, a significant reduction of this uncertainty to below 1\,eV is possible. In the present experiment, the statistical uncertainty even for the most intense K$\alpha$ transition is close to 5\,eV. One reason is that the number of ions injected into the recently commissioned CRYRING@ESR was about an order of magnitude below the design value. In future measurements we expect a significant boost in ion intensity and hence in the counting statistics. Combined with other already planned incremental improvements, such as thicker absorber pixels to achieve a higher quantum efficiency, the presented approach clearly has the potential to reach the 1\,eV accuracy level necessary to test higher-order QED contributions in the heaviest atomic systems.

\begin{acknowledgments}
The authors are indebted to the local teams at GSI, in particular of the ESR and CRYRING@ESR, for making this study possible. We also would like to thank C. E. Düllmann and D. Renisch for providing us with gamma sources for detector calibration. This research has been conducted in the framework of the SPARC collaboration, experiment E138 of FAIR Phase-0 supported by GSI, Darmstadt (Germany). It is further supported by the European Research Council (ERC) under the European Union’s Horizon 2020 research and innovation programme (Grant agreements No. 824109 ``EMP’’ and No. 682841 ``ASTRUm’’). B. Zhu acknowledges CSC Doctoral Fellowship 2018.9 - 2022.2 (Grant No. 201806180051). We also acknowledge the support provided by ErUM FSP T05 - “Aufbau von APPA bei FAIR” (BMBF n° 05P19SJFAA and n° 05P19VHFA1).
\end{acknowledgments}

\bibliography{references}

\begin{thebibliography}{43}%
\makeatletter
\providecommand \@ifxundefined [1]{%
 \@ifx{#1\undefined}
}%
\providecommand \@ifnum [1]{%
 \ifnum #1\expandafter \@firstoftwo
 \else \expandafter \@secondoftwo
 \fi
}%
\providecommand \@ifx [1]{%
 \ifx #1\expandafter \@firstoftwo
 \else \expandafter \@secondoftwo
 \fi
}%
\providecommand \natexlab [1]{#1}%
\providecommand \enquote  [1]{``#1''}%
\providecommand \bibnamefont  [1]{#1}%
\providecommand \bibfnamefont [1]{#1}%
\providecommand \citenamefont [1]{#1}%
\providecommand \href@noop [0]{\@secondoftwo}%
\providecommand \href [0]{\begingroup \@sanitize@url \@href}%
\providecommand \@href[1]{\@@startlink{#1}\@@href}%
\providecommand \@@href[1]{\endgroup#1\@@endlink}%
\providecommand \@sanitize@url [0]{\catcode `\\12\catcode `\$12\catcode
  `\&12\catcode `\#12\catcode `\^12\catcode `\_12\catcode `\%12\relax}%
\providecommand \@@startlink[1]{}%
\providecommand \@@endlink[0]{}%
\providecommand \url  [0]{\begingroup\@sanitize@url \@url }%
\providecommand \@url [1]{\endgroup\@href {#1}{\urlprefix }}%
\providecommand \urlprefix  [0]{URL }%
\providecommand \Eprint [0]{\href }%
\providecommand \doibase [0]{https://doi.org/}%
\providecommand \selectlanguage [0]{\@gobble}%
\providecommand \bibinfo  [0]{\@secondoftwo}%
\providecommand \bibfield  [0]{\@secondoftwo}%
\providecommand \translation [1]{[#1]}%
\providecommand \BibitemOpen [0]{}%
\providecommand \bibitemStop [0]{}%
\providecommand \bibitemNoStop [0]{.\EOS\space}%
\providecommand \EOS [0]{\spacefactor3000\relax}%
\providecommand \BibitemShut  [1]{\csname bibitem#1\endcsname}%
\let\auto@bib@innerbib\@empty
\bibitem [{\citenamefont {Shabaev}\ \emph {et~al.}(2018)\citenamefont
  {Shabaev}, \citenamefont {Bondarev}, \citenamefont {Glazov}, \citenamefont
  {Kaygorodov}, \citenamefont {Kozhedub}, \citenamefont {Maltsev},
  \citenamefont {Malyshev}, \citenamefont {Popov}, \citenamefont {Tupitsyn},\
  and\ \citenamefont {Zubova}}]{Shabaev2018}%
  \BibitemOpen
  \bibfield  {author} {\bibinfo {author} {\bibfnamefont {V.~M.}\ \bibnamefont
  {Shabaev}}, \bibinfo {author} {\bibfnamefont {A.~I.}\ \bibnamefont
  {Bondarev}}, \bibinfo {author} {\bibfnamefont {D.~A.}\ \bibnamefont
  {Glazov}}, \bibinfo {author} {\bibfnamefont {M.~Y.}\ \bibnamefont
  {Kaygorodov}}, \bibinfo {author} {\bibfnamefont {Y.~S.}\ \bibnamefont
  {Kozhedub}}, \bibinfo {author} {\bibfnamefont {I.~A.}\ \bibnamefont
  {Maltsev}}, \bibinfo {author} {\bibfnamefont {A.~V.}\ \bibnamefont
  {Malyshev}}, \bibinfo {author} {\bibfnamefont {R.~V.}\ \bibnamefont {Popov}},
  \bibinfo {author} {\bibfnamefont {I.~I.}\ \bibnamefont {Tupitsyn}},\ and\
  \bibinfo {author} {\bibfnamefont {N.~A.}\ \bibnamefont {Zubova}},\ }\href
  {https://doi.org/10.1007/s10751-018-1537-8} {\bibfield  {journal} {\bibinfo
  {journal} {Hyperfine Interactions}\ }\textbf {\bibinfo {volume} {239}},\
  \bibinfo {pages} {60} (\bibinfo {year} {2018})}\BibitemShut {NoStop}%
\bibitem [{\citenamefont {Yerokhin}\ and\ \citenamefont
  {Shabaev}(2015)}]{Yerokhin2015}%
  \BibitemOpen
  \bibfield  {author} {\bibinfo {author} {\bibfnamefont {V.~A.}\ \bibnamefont
  {Yerokhin}}\ and\ \bibinfo {author} {\bibfnamefont {V.~M.}\ \bibnamefont
  {Shabaev}},\ }\href {https://doi.org/10.1063/1.4927487} {\bibfield  {journal}
  {\bibinfo  {journal} {J. Phys. Chem. Ref. Data}\ }\textbf {\bibinfo {volume}
  {44}},\ \bibinfo {pages} {033103} (\bibinfo {year} {2015})}\BibitemShut
  {NoStop}%
\bibitem [{\citenamefont {Artemyev}\ \emph {et~al.}(2005)\citenamefont
  {Artemyev}, \citenamefont {Shabaev}, \citenamefont {Yerokhin}, \citenamefont
  {Plunien},\ and\ \citenamefont {Soff}}]{Artemyev2005}%
  \BibitemOpen
  \bibfield  {author} {\bibinfo {author} {\bibfnamefont {A.~N.}\ \bibnamefont
  {Artemyev}}, \bibinfo {author} {\bibfnamefont {V.~M.}\ \bibnamefont
  {Shabaev}}, \bibinfo {author} {\bibfnamefont {V.~A.}\ \bibnamefont
  {Yerokhin}}, \bibinfo {author} {\bibfnamefont {G.}~\bibnamefont {Plunien}},\
  and\ \bibinfo {author} {\bibfnamefont {G.}~\bibnamefont {Soff}},\ }\href
  {https://doi.org/10.1103/PhysRevA.71.062104} {\bibfield  {journal} {\bibinfo
  {journal} {Phys. Rev. A}\ }\textbf {\bibinfo {volume} {71}},\ \bibinfo
  {pages} {062104} (\bibinfo {year} {2005})}\BibitemShut {NoStop}%
\bibitem [{\citenamefont {Malyshev}\ \emph {et~al.}(2019)\citenamefont
  {Malyshev}, \citenamefont {Kozhedub}, \citenamefont {Glazov}, \citenamefont
  {Tupitsyn},\ and\ \citenamefont {Shabaev}}]{Malyshev2019}%
  \BibitemOpen
  \bibfield  {author} {\bibinfo {author} {\bibfnamefont {A.~V.}\ \bibnamefont
  {Malyshev}}, \bibinfo {author} {\bibfnamefont {Y.~S.}\ \bibnamefont
  {Kozhedub}}, \bibinfo {author} {\bibfnamefont {D.~A.}\ \bibnamefont
  {Glazov}}, \bibinfo {author} {\bibfnamefont {I.~I.}\ \bibnamefont
  {Tupitsyn}},\ and\ \bibinfo {author} {\bibfnamefont {V.~M.}\ \bibnamefont
  {Shabaev}},\ }\href {https://doi.org/10.1103/PhysRevA.99.010501} {\bibfield
  {journal} {\bibinfo  {journal} {Phys. Rev. A}\ }\textbf {\bibinfo {volume}
  {99}},\ \bibinfo {pages} {010501(R)} (\bibinfo {year} {2019})}\BibitemShut
  {NoStop}%
\bibitem [{\citenamefont {Kozhedub}\ \emph {et~al.}(2019)\citenamefont
  {Kozhedub}, \citenamefont {Malyshev}, \citenamefont {Glazov}, \citenamefont
  {Shabaev},\ and\ \citenamefont {Tupitsyn}}]{Kozhedub2019}%
  \BibitemOpen
  \bibfield  {author} {\bibinfo {author} {\bibfnamefont {Y.~S.}\ \bibnamefont
  {Kozhedub}}, \bibinfo {author} {\bibfnamefont {A.~V.}\ \bibnamefont
  {Malyshev}}, \bibinfo {author} {\bibfnamefont {D.~A.}\ \bibnamefont
  {Glazov}}, \bibinfo {author} {\bibfnamefont {V.~M.}\ \bibnamefont
  {Shabaev}},\ and\ \bibinfo {author} {\bibfnamefont {I.~I.}\ \bibnamefont
  {Tupitsyn}},\ }\href {https://doi.org/10.1103/PhysRevA.100.062506} {\bibfield
   {journal} {\bibinfo  {journal} {Phys. Rev. A}\ }\textbf {\bibinfo {volume}
  {100}},\ \bibinfo {pages} {062506} (\bibinfo {year} {2019})}\BibitemShut
  {NoStop}%
\bibitem [{\citenamefont {Loetzsch}\ \emph {et~al.}(2024)\citenamefont
  {Loetzsch}, \citenamefont {Beyer}, \citenamefont {Duval}, \citenamefont
  {Spillmann}, \citenamefont {Bana{\'{s}}}, \citenamefont {Dergham},
  \citenamefont {Kr{\"o}ger}, \citenamefont {Glorius}, \citenamefont
  {Grisenti}, \citenamefont {Guerra}, \citenamefont {Gumberidze}, \citenamefont
  {He{\ss}}, \citenamefont {Hillenbrand}, \citenamefont {Indelicato},
  \citenamefont {Jagodzinski}, \citenamefont {Lamour}, \citenamefont {Lorentz},
  \citenamefont {Litvinov}, \citenamefont {Litvinov}, \citenamefont {Machado},
  \citenamefont {Paul}, \citenamefont {Paulus}, \citenamefont {Petridis},
  \citenamefont {Santos}, \citenamefont {Scheidel}, \citenamefont {Sidhu},
  \citenamefont {Steck}, \citenamefont {Steydli}, \citenamefont {Szary},
  \citenamefont {Trotsenko}, \citenamefont {Uschmann}, \citenamefont {Weber},
  \citenamefont {St{\"o}hlker},\ and\ \citenamefont
  {Trassinelli}}]{Loetzsch2024}%
  \BibitemOpen
  \bibfield  {author} {\bibinfo {author} {\bibfnamefont {R.}~\bibnamefont
  {Loetzsch}}, \bibinfo {author} {\bibfnamefont {H.~F.}\ \bibnamefont {Beyer}},
  \bibinfo {author} {\bibfnamefont {L.}~\bibnamefont {Duval}}, \bibinfo
  {author} {\bibfnamefont {U.}~\bibnamefont {Spillmann}}, \bibinfo {author}
  {\bibfnamefont {D.}~\bibnamefont {Bana{\'{s}}}}, \bibinfo {author}
  {\bibfnamefont {P.}~\bibnamefont {Dergham}}, \bibinfo {author} {\bibfnamefont
  {F.~M.}\ \bibnamefont {Kr{\"o}ger}}, \bibinfo {author} {\bibfnamefont
  {J.}~\bibnamefont {Glorius}}, \bibinfo {author} {\bibfnamefont {R.~E.}\
  \bibnamefont {Grisenti}}, \bibinfo {author} {\bibfnamefont {M.}~\bibnamefont
  {Guerra}}, \bibinfo {author} {\bibfnamefont {A.}~\bibnamefont {Gumberidze}},
  \bibinfo {author} {\bibfnamefont {R.}~\bibnamefont {He{\ss}}}, \bibinfo
  {author} {\bibfnamefont {P.-M.}\ \bibnamefont {Hillenbrand}}, \bibinfo
  {author} {\bibfnamefont {P.}~\bibnamefont {Indelicato}}, \bibinfo {author}
  {\bibfnamefont {P.}~\bibnamefont {Jagodzinski}}, \bibinfo {author}
  {\bibfnamefont {E.}~\bibnamefont {Lamour}}, \bibinfo {author} {\bibfnamefont
  {B.}~\bibnamefont {Lorentz}}, \bibinfo {author} {\bibfnamefont
  {S.}~\bibnamefont {Litvinov}}, \bibinfo {author} {\bibfnamefont {Y.~A.}\
  \bibnamefont {Litvinov}}, \bibinfo {author} {\bibfnamefont {J.}~\bibnamefont
  {Machado}}, \bibinfo {author} {\bibfnamefont {N.}~\bibnamefont {Paul}},
  \bibinfo {author} {\bibfnamefont {G.~G.}\ \bibnamefont {Paulus}}, \bibinfo
  {author} {\bibfnamefont {N.}~\bibnamefont {Petridis}}, \bibinfo {author}
  {\bibfnamefont {J.~P.}\ \bibnamefont {Santos}}, \bibinfo {author}
  {\bibfnamefont {M.}~\bibnamefont {Scheidel}}, \bibinfo {author}
  {\bibfnamefont {R.~S.}\ \bibnamefont {Sidhu}}, \bibinfo {author}
  {\bibfnamefont {M.}~\bibnamefont {Steck}}, \bibinfo {author} {\bibfnamefont
  {S.}~\bibnamefont {Steydli}}, \bibinfo {author} {\bibfnamefont
  {K.}~\bibnamefont {Szary}}, \bibinfo {author} {\bibfnamefont
  {S.}~\bibnamefont {Trotsenko}}, \bibinfo {author} {\bibfnamefont
  {I.}~\bibnamefont {Uschmann}}, \bibinfo {author} {\bibfnamefont
  {G.}~\bibnamefont {Weber}}, \bibinfo {author} {\bibfnamefont
  {T.}~\bibnamefont {St{\"o}hlker}},\ and\ \bibinfo {author} {\bibfnamefont
  {M.}~\bibnamefont {Trassinelli}},\ }\href
  {https://doi.org/10.1038/s41586-023-06910-y} {\bibfield  {journal} {\bibinfo
  {journal} {Nature}\ }\textbf {\bibinfo {volume} {625}},\ \bibinfo {pages}
  {673} (\bibinfo {year} {2024})}\BibitemShut {NoStop}%
\bibitem [{\citenamefont {Ullmann}\ \emph {et~al.}(2017)\citenamefont
  {Ullmann}, \citenamefont {Andelkovic}, \citenamefont {Brandau}, \citenamefont
  {Dax}, \citenamefont {Geithner}, \citenamefont {Geppert}, \citenamefont
  {Gorges}, \citenamefont {Hammen}, \citenamefont {Hannen}, \citenamefont
  {Kaufmann}, \citenamefont {K\"onig}, \citenamefont {Litvinov}, \citenamefont
  {Lochmann}, \citenamefont {Maaß}, \citenamefont {Meisner}, \citenamefont
  {Murb\"ock}, \citenamefont {Sánchez}, \citenamefont {Schmidt}, \citenamefont
  {Schmidt}, \citenamefont {Steck}, \citenamefont {St\"ohlker}, \citenamefont
  {Thompson}, \citenamefont {Trageser}, \citenamefont {Vollbrecht},
  \citenamefont {Weinheimer},\ and\ \citenamefont
  {N\"ortersh\"auser}}]{ullmann2017}%
  \BibitemOpen
  \bibfield  {author} {\bibinfo {author} {\bibfnamefont {J.}~\bibnamefont
  {Ullmann}}, \bibinfo {author} {\bibfnamefont {Z.}~\bibnamefont {Andelkovic}},
  \bibinfo {author} {\bibfnamefont {C.}~\bibnamefont {Brandau}}, \bibinfo
  {author} {\bibfnamefont {A.}~\bibnamefont {Dax}}, \bibinfo {author}
  {\bibfnamefont {W.}~\bibnamefont {Geithner}}, \bibinfo {author}
  {\bibfnamefont {C.}~\bibnamefont {Geppert}}, \bibinfo {author} {\bibfnamefont
  {C.}~\bibnamefont {Gorges}}, \bibinfo {author} {\bibfnamefont
  {M.}~\bibnamefont {Hammen}}, \bibinfo {author} {\bibfnamefont
  {V.}~\bibnamefont {Hannen}}, \bibinfo {author} {\bibfnamefont
  {S.}~\bibnamefont {Kaufmann}}, \bibinfo {author} {\bibfnamefont
  {K.}~\bibnamefont {K\"onig}}, \bibinfo {author} {\bibfnamefont {Y.~A.}\
  \bibnamefont {Litvinov}}, \bibinfo {author} {\bibfnamefont {M.}~\bibnamefont
  {Lochmann}}, \bibinfo {author} {\bibfnamefont {B.}~\bibnamefont {Maaß}},
  \bibinfo {author} {\bibfnamefont {J.}~\bibnamefont {Meisner}}, \bibinfo
  {author} {\bibfnamefont {T.}~\bibnamefont {Murb\"ock}}, \bibinfo {author}
  {\bibfnamefont {R.}~\bibnamefont {Sánchez}}, \bibinfo {author}
  {\bibfnamefont {M.}~\bibnamefont {Schmidt}}, \bibinfo {author} {\bibfnamefont
  {S.}~\bibnamefont {Schmidt}}, \bibinfo {author} {\bibfnamefont
  {M.}~\bibnamefont {Steck}}, \bibinfo {author} {\bibfnamefont {{\relax
  Th}.}~\bibnamefont {St\"ohlker}}, \bibinfo {author} {\bibfnamefont {R.~C.}\
  \bibnamefont {Thompson}}, \bibinfo {author} {\bibfnamefont {C.}~\bibnamefont
  {Trageser}}, \bibinfo {author} {\bibfnamefont {J.}~\bibnamefont
  {Vollbrecht}}, \bibinfo {author} {\bibfnamefont {C.}~\bibnamefont
  {Weinheimer}},\ and\ \bibinfo {author} {\bibfnamefont {W.}~\bibnamefont
  {N\"ortersh\"auser}},\ }\href {https://doi.org/10.1038/ncomms15484}
  {\bibfield  {journal} {\bibinfo  {journal} {Nat Commun}\ }\textbf {\bibinfo
  {volume} {8}},\ \bibinfo {pages} {15484} (\bibinfo {year}
  {2017})}\BibitemShut {NoStop}%
\bibitem [{\citenamefont {Skripnikov}\ \emph {et~al.}(2018)\citenamefont
  {Skripnikov}, \citenamefont {Schmidt}, \citenamefont {Ullmann}, \citenamefont
  {Geppert}, \citenamefont {Kraus}, \citenamefont {Kresse}, \citenamefont
  {N\"ortersh\"auser}, \citenamefont {Privalov}, \citenamefont {Scheibe},
  \citenamefont {Shabaev}, \citenamefont {Vogel},\ and\ \citenamefont
  {Volotka}}]{skripnikov2018}%
  \BibitemOpen
  \bibfield  {author} {\bibinfo {author} {\bibfnamefont {L.~V.}\ \bibnamefont
  {Skripnikov}}, \bibinfo {author} {\bibfnamefont {S.}~\bibnamefont {Schmidt}},
  \bibinfo {author} {\bibfnamefont {J.}~\bibnamefont {Ullmann}}, \bibinfo
  {author} {\bibfnamefont {C.}~\bibnamefont {Geppert}}, \bibinfo {author}
  {\bibfnamefont {F.}~\bibnamefont {Kraus}}, \bibinfo {author} {\bibfnamefont
  {B.}~\bibnamefont {Kresse}}, \bibinfo {author} {\bibfnamefont
  {W.}~\bibnamefont {N\"ortersh\"auser}}, \bibinfo {author} {\bibfnamefont
  {A.~F.}\ \bibnamefont {Privalov}}, \bibinfo {author} {\bibfnamefont
  {B.}~\bibnamefont {Scheibe}}, \bibinfo {author} {\bibfnamefont {V.~M.}\
  \bibnamefont {Shabaev}}, \bibinfo {author} {\bibfnamefont {M.}~\bibnamefont
  {Vogel}},\ and\ \bibinfo {author} {\bibfnamefont {A.~V.}\ \bibnamefont
  {Volotka}},\ }\href {https://doi.org/10.1103/PhysRevLett.120.093001}
  {\bibfield  {journal} {\bibinfo  {journal} {Phys. Rev. Lett.}\ }\textbf
  {\bibinfo {volume} {120}},\ \bibinfo {pages} {093001} (\bibinfo {year}
  {2018})}\BibitemShut {NoStop}%
\bibitem [{\citenamefont {Morgner}\ \emph {et~al.}(2023)\citenamefont
  {Morgner}, \citenamefont {Tu}, \citenamefont {K\"onig}, \citenamefont
  {Sailer}, \citenamefont {Heiße}, \citenamefont {Bekker}, \citenamefont
  {Sikora}, \citenamefont {Lyu}, \citenamefont {Yerokhin}, \citenamefont
  {Harman}, \citenamefont {Crespo López-Urrutia}, \citenamefont {Keitel},
  \citenamefont {Sturm},\ and\ \citenamefont {Blaum}}]{Morgner2023}%
  \BibitemOpen
  \bibfield  {author} {\bibinfo {author} {\bibfnamefont {J.}~\bibnamefont
  {Morgner}}, \bibinfo {author} {\bibfnamefont {B.}~\bibnamefont {Tu}},
  \bibinfo {author} {\bibfnamefont {C.~M.}\ \bibnamefont {K\"onig}}, \bibinfo
  {author} {\bibfnamefont {T.}~\bibnamefont {Sailer}}, \bibinfo {author}
  {\bibfnamefont {F.}~\bibnamefont {Heiße}}, \bibinfo {author} {\bibfnamefont
  {H.}~\bibnamefont {Bekker}}, \bibinfo {author} {\bibfnamefont
  {B.}~\bibnamefont {Sikora}}, \bibinfo {author} {\bibfnamefont
  {C.}~\bibnamefont {Lyu}}, \bibinfo {author} {\bibfnamefont {V.~A.}\
  \bibnamefont {Yerokhin}}, \bibinfo {author} {\bibfnamefont {Z.}~\bibnamefont
  {Harman}}, \bibinfo {author} {\bibfnamefont {J.~R.}\ \bibnamefont {Crespo
  López-Urrutia}}, \bibinfo {author} {\bibfnamefont {C.~H.}\ \bibnamefont
  {Keitel}}, \bibinfo {author} {\bibfnamefont {S.}~\bibnamefont {Sturm}},\ and\
  \bibinfo {author} {\bibfnamefont {K.}~\bibnamefont {Blaum}},\ }\href
  {https://doi.org/10.1038/s41586-023-06453-2} {\bibfield  {journal} {\bibinfo
  {journal} {Nature}\ }\textbf {\bibinfo {volume} {622}},\ \bibinfo {pages}
  {53} (\bibinfo {year} {2023})}\BibitemShut {NoStop}%
\bibitem [{\citenamefont {Beiersdorfer}(2010)}]{Beiersdorfer2010}%
  \BibitemOpen
  \bibfield  {author} {\bibinfo {author} {\bibfnamefont {P.}~\bibnamefont
  {Beiersdorfer}},\ }\href {https://doi.org/10.1088/0953-4075/43/7/074032}
  {\bibfield  {journal} {\bibinfo  {journal} {Journal of Physics B: Atomic,
  Molecular and Optical Physics}\ }\textbf {\bibinfo {volume} {43}},\ \bibinfo
  {pages} {074032} (\bibinfo {year} {2010})}\BibitemShut {NoStop}%
\bibitem [{\citenamefont {Indelicato}(2019)}]{Indelicato2019}%
  \BibitemOpen
  \bibfield  {author} {\bibinfo {author} {\bibfnamefont {P.}~\bibnamefont
  {Indelicato}},\ }\href {https://doi.org/10.1088/1361-6455/ab42c9} {\bibfield
  {journal} {\bibinfo  {journal} {Journal of Physics B: Atomic, Molecular and
  Optical Physics}\ }\textbf {\bibinfo {volume} {52}},\ \bibinfo {pages}
  {232001} (\bibinfo {year} {2019})}\BibitemShut {NoStop}%
\bibitem [{\citenamefont {Kozlov}\ \emph {et~al.}(2018)\citenamefont {Kozlov},
  \citenamefont {Safronova}, \citenamefont {Crespo L\'{o}pez-Urrutia},\ and\
  \citenamefont {Schmidt}}]{kozlov_highly_2018}%
  \BibitemOpen
  \bibfield  {author} {\bibinfo {author} {\bibfnamefont {M.~G.}\ \bibnamefont
  {Kozlov}}, \bibinfo {author} {\bibfnamefont {M.~S.}\ \bibnamefont
  {Safronova}}, \bibinfo {author} {\bibfnamefont {J.~R.}\ \bibnamefont {Crespo
  L\'{o}pez-Urrutia}},\ and\ \bibinfo {author} {\bibfnamefont {P.~O.}\
  \bibnamefont {Schmidt}},\ }\href
  {https://doi.org/10.1103/RevModPhys.90.045005} {\bibfield  {journal}
  {\bibinfo  {journal} {Rev. Mod. Phys.}\ }\textbf {\bibinfo {volume} {90}},\
  \bibinfo {pages} {045005} (\bibinfo {year} {2018})}\BibitemShut {NoStop}%
\bibitem [{\citenamefont {Safronova}\ \emph {et~al.}(2018)\citenamefont
  {Safronova}, \citenamefont {Budker}, \citenamefont {DeMille}, \citenamefont
  {Kimball}, \citenamefont {Derevianko},\ and\ \citenamefont
  {Clark}}]{safronova_search_2018}%
  \BibitemOpen
  \bibfield  {author} {\bibinfo {author} {\bibfnamefont {M.~S.}\ \bibnamefont
  {Safronova}}, \bibinfo {author} {\bibfnamefont {D.}~\bibnamefont {Budker}},
  \bibinfo {author} {\bibfnamefont {D.}~\bibnamefont {DeMille}}, \bibinfo
  {author} {\bibfnamefont {D.~F.~J.}\ \bibnamefont {Kimball}}, \bibinfo
  {author} {\bibfnamefont {A.}~\bibnamefont {Derevianko}},\ and\ \bibinfo
  {author} {\bibfnamefont {C.~W.}\ \bibnamefont {Clark}},\ }\href
  {https://doi.org/10.1103/RevModPhys.90.025008} {\bibfield  {journal}
  {\bibinfo  {journal} {Rev. Mod. Phys.}\ }\textbf {\bibinfo {volume} {90}},\
  \bibinfo {pages} {025008} (\bibinfo {year} {2018})}\BibitemShut {NoStop}%
\bibitem [{\citenamefont {Safronova}(2019)}]{Safronova2019}%
  \BibitemOpen
  \bibfield  {author} {\bibinfo {author} {\bibfnamefont {M.~S.}\ \bibnamefont
  {Safronova}},\ }\href
  {https://doi.org/https://doi.org/10.1002/andp.201800364} {\bibfield
  {journal} {\bibinfo  {journal} {Annalen der Physik}\ }\textbf {\bibinfo
  {volume} {531}},\ \bibinfo {pages} {1800364} (\bibinfo {year}
  {2019})}\BibitemShut {NoStop}%
\bibitem [{\citenamefont {Gabriel}(1972)}]{Gabriel1972}%
  \BibitemOpen
  \bibfield  {author} {\bibinfo {author} {\bibfnamefont {A.~H.}\ \bibnamefont
  {Gabriel}},\ }\href {https://doi.org/10.1093/mnras/160.1.99} {\bibfield
  {journal} {\bibinfo  {journal} {Monthly Notices of the Royal Astronomical
  Society}\ }\textbf {\bibinfo {volume} {160}},\ \bibinfo {pages} {99}
  (\bibinfo {year} {1972})}\BibitemShut {NoStop}%
\bibitem [{\citenamefont {Thorn}\ \emph {et~al.}(2009)\citenamefont {Thorn},
  \citenamefont {Gu}, \citenamefont {Brown}, \citenamefont {Beiersdorfer},
  \citenamefont {Porter}, \citenamefont {Kilbourne},\ and\ \citenamefont
  {Kelley}}]{Thorn2009}%
  \BibitemOpen
  \bibfield  {author} {\bibinfo {author} {\bibfnamefont {D.~B.}\ \bibnamefont
  {Thorn}}, \bibinfo {author} {\bibfnamefont {M.~F.}\ \bibnamefont {Gu}},
  \bibinfo {author} {\bibfnamefont {G.~V.}\ \bibnamefont {Brown}}, \bibinfo
  {author} {\bibfnamefont {P.}~\bibnamefont {Beiersdorfer}}, \bibinfo {author}
  {\bibfnamefont {F.~S.}\ \bibnamefont {Porter}}, \bibinfo {author}
  {\bibfnamefont {C.~A.}\ \bibnamefont {Kilbourne}},\ and\ \bibinfo {author}
  {\bibfnamefont {R.~L.}\ \bibnamefont {Kelley}},\ }\href
  {https://doi.org/10.1103/PhysRevLett.103.163001} {\bibfield  {journal}
  {\bibinfo  {journal} {Phys. Rev. Lett.}\ }\textbf {\bibinfo {volume} {103}},\
  \bibinfo {pages} {163001} (\bibinfo {year} {2009})}\BibitemShut {NoStop}%
\bibitem [{\citenamefont {Chantler}\ \emph {et~al.}(2012)\citenamefont
  {Chantler}, \citenamefont {Kinnane}, \citenamefont {Gillaspy}, \citenamefont
  {Hudson}, \citenamefont {Payne}, \citenamefont {Smale}, \citenamefont
  {Henins}, \citenamefont {Pomeroy}, \citenamefont {Tan}, \citenamefont
  {Kimpton}, \citenamefont {Takacs},\ and\ \citenamefont
  {Makonyi}}]{Chantler2012}%
  \BibitemOpen
  \bibfield  {author} {\bibinfo {author} {\bibfnamefont {C.~T.}\ \bibnamefont
  {Chantler}}, \bibinfo {author} {\bibfnamefont {M.~N.}\ \bibnamefont
  {Kinnane}}, \bibinfo {author} {\bibfnamefont {J.~D.}\ \bibnamefont
  {Gillaspy}}, \bibinfo {author} {\bibfnamefont {L.~T.}\ \bibnamefont
  {Hudson}}, \bibinfo {author} {\bibfnamefont {A.~T.}\ \bibnamefont {Payne}},
  \bibinfo {author} {\bibfnamefont {L.~F.}\ \bibnamefont {Smale}}, \bibinfo
  {author} {\bibfnamefont {A.}~\bibnamefont {Henins}}, \bibinfo {author}
  {\bibfnamefont {J.~M.}\ \bibnamefont {Pomeroy}}, \bibinfo {author}
  {\bibfnamefont {J.~N.}\ \bibnamefont {Tan}}, \bibinfo {author} {\bibfnamefont
  {J.~A.}\ \bibnamefont {Kimpton}}, \bibinfo {author} {\bibfnamefont
  {E.}~\bibnamefont {Takacs}},\ and\ \bibinfo {author} {\bibfnamefont
  {K.}~\bibnamefont {Makonyi}},\ }\href
  {https://doi.org/10.1103/PhysRevLett.109.153001} {\bibfield  {journal}
  {\bibinfo  {journal} {Phys. Rev. Lett.}\ }\textbf {\bibinfo {volume} {109}},\
  \bibinfo {pages} {153001} (\bibinfo {year} {2012})}\BibitemShut {NoStop}%
\bibitem [{\citenamefont {Epp}(2013)}]{Epp2013}%
  \BibitemOpen
  \bibfield  {author} {\bibinfo {author} {\bibfnamefont {S.~W.}\ \bibnamefont
  {Epp}},\ }\href {https://doi.org/10.1103/PhysRevLett.110.159301} {\bibfield
  {journal} {\bibinfo  {journal} {Phys. Rev. Lett.}\ }\textbf {\bibinfo
  {volume} {110}},\ \bibinfo {pages} {159301} (\bibinfo {year}
  {2013})}\BibitemShut {NoStop}%
\bibitem [{\citenamefont {Chantler}\ \emph {et~al.}(2013)\citenamefont
  {Chantler}, \citenamefont {Kinnane}, \citenamefont {Gillaspy}, \citenamefont
  {Hudson}, \citenamefont {Payne}, \citenamefont {Smale}, \citenamefont
  {Henins}, \citenamefont {Pomeroy}, \citenamefont {Kimpton}, \citenamefont
  {Takacs},\ and\ \citenamefont {Makonyi}}]{Chantler2013}%
  \BibitemOpen
  \bibfield  {author} {\bibinfo {author} {\bibfnamefont {C.~T.}\ \bibnamefont
  {Chantler}}, \bibinfo {author} {\bibfnamefont {M.~N.}\ \bibnamefont
  {Kinnane}}, \bibinfo {author} {\bibfnamefont {J.~D.}\ \bibnamefont
  {Gillaspy}}, \bibinfo {author} {\bibfnamefont {L.~T.}\ \bibnamefont
  {Hudson}}, \bibinfo {author} {\bibfnamefont {A.~T.}\ \bibnamefont {Payne}},
  \bibinfo {author} {\bibfnamefont {L.~F.}\ \bibnamefont {Smale}}, \bibinfo
  {author} {\bibfnamefont {A.}~\bibnamefont {Henins}}, \bibinfo {author}
  {\bibfnamefont {J.~M.}\ \bibnamefont {Pomeroy}}, \bibinfo {author}
  {\bibfnamefont {J.~A.}\ \bibnamefont {Kimpton}}, \bibinfo {author}
  {\bibfnamefont {E.}~\bibnamefont {Takacs}},\ and\ \bibinfo {author}
  {\bibfnamefont {K.}~\bibnamefont {Makonyi}},\ }\href
  {https://doi.org/10.1103/PhysRevLett.110.159302} {\bibfield  {journal}
  {\bibinfo  {journal} {Phys. Rev. Lett.}\ }\textbf {\bibinfo {volume} {110}},\
  \bibinfo {pages} {159302} (\bibinfo {year} {2013})}\BibitemShut {NoStop}%
\bibitem [{\citenamefont {Payne}\ \emph {et~al.}(2014)\citenamefont {Payne},
  \citenamefont {Chantler}, \citenamefont {Kinnane}, \citenamefont {Gillaspy},
  \citenamefont {Hudson}, \citenamefont {Smale}, \citenamefont {Henins},
  \citenamefont {Kimpton},\ and\ \citenamefont {Takacs}}]{Payne2014}%
  \BibitemOpen
  \bibfield  {author} {\bibinfo {author} {\bibfnamefont {A.~T.}\ \bibnamefont
  {Payne}}, \bibinfo {author} {\bibfnamefont {C.~T.}\ \bibnamefont {Chantler}},
  \bibinfo {author} {\bibfnamefont {M.~N.}\ \bibnamefont {Kinnane}}, \bibinfo
  {author} {\bibfnamefont {J.~D.}\ \bibnamefont {Gillaspy}}, \bibinfo {author}
  {\bibfnamefont {L.~T.}\ \bibnamefont {Hudson}}, \bibinfo {author}
  {\bibfnamefont {L.~F.}\ \bibnamefont {Smale}}, \bibinfo {author}
  {\bibfnamefont {A.}~\bibnamefont {Henins}}, \bibinfo {author} {\bibfnamefont
  {J.~A.}\ \bibnamefont {Kimpton}},\ and\ \bibinfo {author} {\bibfnamefont
  {E.}~\bibnamefont {Takacs}},\ }\href
  {https://doi.org/10.1088/0953-4075/47/18/185001} {\bibfield  {journal}
  {\bibinfo  {journal} {Journal of Physics B: Atomic, Molecular and Optical
  Physics}\ }\textbf {\bibinfo {volume} {47}},\ \bibinfo {pages} {185001}
  (\bibinfo {year} {2014})}\BibitemShut {NoStop}%
\bibitem [{\citenamefont {Kubi\ifmmode~\check{c}\else \v{c}\fi{}ek}\ \emph
  {et~al.}(2014)\citenamefont {Kubi\ifmmode~\check{c}\else \v{c}\fi{}ek},
  \citenamefont {Mokler}, \citenamefont {M\"ackel}, \citenamefont {Ullrich},\
  and\ \citenamefont {Crespo L\'opez-Urrutia}}]{Kubicek2014}%
  \BibitemOpen
  \bibfield  {author} {\bibinfo {author} {\bibfnamefont {K.}~\bibnamefont
  {Kubi\ifmmode~\check{c}\else \v{c}\fi{}ek}}, \bibinfo {author} {\bibfnamefont
  {P.~H.}\ \bibnamefont {Mokler}}, \bibinfo {author} {\bibfnamefont
  {V.}~\bibnamefont {M\"ackel}}, \bibinfo {author} {\bibfnamefont
  {J.}~\bibnamefont {Ullrich}},\ and\ \bibinfo {author} {\bibfnamefont {J.~R.}\
  \bibnamefont {Crespo L\'opez-Urrutia}},\ }\href
  {https://doi.org/10.1103/PhysRevA.90.032508} {\bibfield  {journal} {\bibinfo
  {journal} {Phys. Rev. A}\ }\textbf {\bibinfo {volume} {90}},\ \bibinfo
  {pages} {032508} (\bibinfo {year} {2014})}\BibitemShut {NoStop}%
\bibitem [{\citenamefont {Chantler}\ \emph {et~al.}(2014)\citenamefont
  {Chantler}, \citenamefont {Payne}, \citenamefont {Gillaspy}, \citenamefont
  {Hudson}, \citenamefont {Smale}, \citenamefont {Henins}, \citenamefont
  {Kimpton},\ and\ \citenamefont {Takacs}}]{Chantler2014}%
  \BibitemOpen
  \bibfield  {author} {\bibinfo {author} {\bibfnamefont {C.~T.}\ \bibnamefont
  {Chantler}}, \bibinfo {author} {\bibfnamefont {A.~T.}\ \bibnamefont {Payne}},
  \bibinfo {author} {\bibfnamefont {J.~D.}\ \bibnamefont {Gillaspy}}, \bibinfo
  {author} {\bibfnamefont {L.~T.}\ \bibnamefont {Hudson}}, \bibinfo {author}
  {\bibfnamefont {L.~F.}\ \bibnamefont {Smale}}, \bibinfo {author}
  {\bibfnamefont {A.}~\bibnamefont {Henins}}, \bibinfo {author} {\bibfnamefont
  {J.~A.}\ \bibnamefont {Kimpton}},\ and\ \bibinfo {author} {\bibfnamefont
  {E.}~\bibnamefont {Takacs}},\ }\href
  {https://doi.org/10.1088/1367-2630/16/12/123037} {\bibfield  {journal}
  {\bibinfo  {journal} {New J. Phys.}\ }\textbf {\bibinfo {volume} {16}},\
  \bibinfo {pages} {123037} (\bibinfo {year} {2014})}\BibitemShut {NoStop}%
\bibitem [{\citenamefont {Epp}\ \emph {et~al.}(2015)\citenamefont {Epp},
  \citenamefont {Steinbr\"ugge}, \citenamefont {Bernitt}, \citenamefont
  {Rudolph}, \citenamefont {Beilmann}, \citenamefont {Bekker}, \citenamefont
  {M\"uller}, \citenamefont {Versolato}, \citenamefont {Wille}, \citenamefont
  {Yava\ifmmode~\mbox{\c{s}}\else \c{s}\fi{}}, \citenamefont {Ullrich},\ and\
  \citenamefont {Crespo L\'opez-Urrutia}}]{Epp2015}%
  \BibitemOpen
  \bibfield  {author} {\bibinfo {author} {\bibfnamefont {S.~W.}\ \bibnamefont
  {Epp}}, \bibinfo {author} {\bibfnamefont {R.}~\bibnamefont {Steinbr\"ugge}},
  \bibinfo {author} {\bibfnamefont {S.}~\bibnamefont {Bernitt}}, \bibinfo
  {author} {\bibfnamefont {J.~K.}\ \bibnamefont {Rudolph}}, \bibinfo {author}
  {\bibfnamefont {C.}~\bibnamefont {Beilmann}}, \bibinfo {author}
  {\bibfnamefont {H.}~\bibnamefont {Bekker}}, \bibinfo {author} {\bibfnamefont
  {A.}~\bibnamefont {M\"uller}}, \bibinfo {author} {\bibfnamefont {O.~O.}\
  \bibnamefont {Versolato}}, \bibinfo {author} {\bibfnamefont {H.-C.}\
  \bibnamefont {Wille}}, \bibinfo {author} {\bibfnamefont {H.}~\bibnamefont
  {Yava\ifmmode~\mbox{\c{s}}\else \c{s}\fi{}}}, \bibinfo {author}
  {\bibfnamefont {J.}~\bibnamefont {Ullrich}},\ and\ \bibinfo {author}
  {\bibfnamefont {J.~R.}\ \bibnamefont {Crespo L\'opez-Urrutia}},\ }\href
  {https://doi.org/10.1103/PhysRevA.92.020502} {\bibfield  {journal} {\bibinfo
  {journal} {Phys. Rev. A}\ }\textbf {\bibinfo {volume} {92}},\ \bibinfo
  {pages} {020502(R)} (\bibinfo {year} {2015})}\BibitemShut {NoStop}%
\bibitem [{\citenamefont {Beiersdorfer}\ and\ \citenamefont
  {Brown}(2015)}]{Beiersdorfer2015}%
  \BibitemOpen
  \bibfield  {author} {\bibinfo {author} {\bibfnamefont {P.}~\bibnamefont
  {Beiersdorfer}}\ and\ \bibinfo {author} {\bibfnamefont {G.~V.}\ \bibnamefont
  {Brown}},\ }\href {https://doi.org/10.1103/PhysRevA.91.032514} {\bibfield
  {journal} {\bibinfo  {journal} {Phys. Rev. A}\ }\textbf {\bibinfo {volume}
  {91}},\ \bibinfo {pages} {032514} (\bibinfo {year} {2015})}\BibitemShut
  {NoStop}%
\bibitem [{\citenamefont {Machado}\ \emph {et~al.}(2018)\citenamefont
  {Machado}, \citenamefont {Szabo}, \citenamefont {Santos}, \citenamefont
  {Amaro}, \citenamefont {Guerra}, \citenamefont {Gumberidze}, \citenamefont
  {Bian}, \citenamefont {Isac},\ and\ \citenamefont
  {Indelicato}}]{Machado2018}%
  \BibitemOpen
  \bibfield  {author} {\bibinfo {author} {\bibfnamefont {J.}~\bibnamefont
  {Machado}}, \bibinfo {author} {\bibfnamefont {C.~I.}\ \bibnamefont {Szabo}},
  \bibinfo {author} {\bibfnamefont {J.~P.}\ \bibnamefont {Santos}}, \bibinfo
  {author} {\bibfnamefont {P.}~\bibnamefont {Amaro}}, \bibinfo {author}
  {\bibfnamefont {M.}~\bibnamefont {Guerra}}, \bibinfo {author} {\bibfnamefont
  {A.}~\bibnamefont {Gumberidze}}, \bibinfo {author} {\bibfnamefont
  {G.}~\bibnamefont {Bian}}, \bibinfo {author} {\bibfnamefont {J.~M.}\
  \bibnamefont {Isac}},\ and\ \bibinfo {author} {\bibfnamefont
  {P.}~\bibnamefont {Indelicato}},\ }\href
  {https://doi.org/10.1103/PhysRevA.97.032517} {\bibfield  {journal} {\bibinfo
  {journal} {Phys. Rev. A}\ }\textbf {\bibinfo {volume} {97}},\ \bibinfo
  {pages} {032517} (\bibinfo {year} {2018})}\BibitemShut {NoStop}%
\bibitem [{\citenamefont {Briand}\ \emph {et~al.}(1990)\citenamefont {Briand},
  \citenamefont {Chevallier}, \citenamefont {Indelicato}, \citenamefont
  {Ziock},\ and\ \citenamefont {Dietrich}}]{Briand90}%
  \BibitemOpen
  \bibfield  {author} {\bibinfo {author} {\bibfnamefont {J.~P.}\ \bibnamefont
  {Briand}}, \bibinfo {author} {\bibfnamefont {P.}~\bibnamefont {Chevallier}},
  \bibinfo {author} {\bibfnamefont {P.}~\bibnamefont {Indelicato}}, \bibinfo
  {author} {\bibfnamefont {K.~P.}\ \bibnamefont {Ziock}},\ and\ \bibinfo
  {author} {\bibfnamefont {D.~D.}\ \bibnamefont {Dietrich}},\ }\href
  {https://doi.org/10.1103/PhysRevLett.65.2761} {\bibfield  {journal} {\bibinfo
   {journal} {Phys. Rev. Lett.}\ }\textbf {\bibinfo {volume} {65}},\ \bibinfo
  {pages} {2761} (\bibinfo {year} {1990})}\BibitemShut {NoStop}%
\bibitem [{\citenamefont {Gumberidze}\ \emph {et~al.}(2004)\citenamefont
  {Gumberidze}, \citenamefont {St\"ohlker}, \citenamefont
  {Bana\ifmmode~\acute{s}\else \'{s}\fi{}}, \citenamefont {Beckert},
  \citenamefont {Beller}, \citenamefont {Beyer}, \citenamefont {Bosch},
  \citenamefont {Cai}, \citenamefont {Hagmann}, \citenamefont {Kozhuharov},
  \citenamefont {Liesen}, \citenamefont {Nolden}, \citenamefont {Ma},
  \citenamefont {Mokler}, \citenamefont {Or\ifmmode \check{s}\else
  \v{s}\fi{}i\ifmmode \acute{c}\else~\'{c}\fi{} Muthig}, \citenamefont {Steck},
  \citenamefont {Sierpowski}, \citenamefont {Tashenov}, \citenamefont
  {Warczak},\ and\ \citenamefont {Zou}}]{Gumberidze2004}%
  \BibitemOpen
  \bibfield  {author} {\bibinfo {author} {\bibfnamefont {A.}~\bibnamefont
  {Gumberidze}}, \bibinfo {author} {\bibfnamefont {{\relax Th}.}~\bibnamefont
  {St\"ohlker}}, \bibinfo {author} {\bibfnamefont {D.}~\bibnamefont
  {Bana\ifmmode~\acute{s}\else \'{s}\fi{}}}, \bibinfo {author} {\bibfnamefont
  {K.}~\bibnamefont {Beckert}}, \bibinfo {author} {\bibfnamefont
  {P.}~\bibnamefont {Beller}}, \bibinfo {author} {\bibfnamefont {H.~F.}\
  \bibnamefont {Beyer}}, \bibinfo {author} {\bibfnamefont {F.}~\bibnamefont
  {Bosch}}, \bibinfo {author} {\bibfnamefont {X.}~\bibnamefont {Cai}}, \bibinfo
  {author} {\bibfnamefont {S.}~\bibnamefont {Hagmann}}, \bibinfo {author}
  {\bibfnamefont {C.}~\bibnamefont {Kozhuharov}}, \bibinfo {author}
  {\bibfnamefont {D.}~\bibnamefont {Liesen}}, \bibinfo {author} {\bibfnamefont
  {F.}~\bibnamefont {Nolden}}, \bibinfo {author} {\bibfnamefont
  {X.}~\bibnamefont {Ma}}, \bibinfo {author} {\bibfnamefont {P.~H.}\
  \bibnamefont {Mokler}}, \bibinfo {author} {\bibfnamefont {A.}~\bibnamefont
  {Or\ifmmode \check{s}\else \v{s}\fi{}i\ifmmode \acute{c}\else~\'{c}\fi{}
  Muthig}}, \bibinfo {author} {\bibfnamefont {M.}~\bibnamefont {Steck}},
  \bibinfo {author} {\bibfnamefont {D.}~\bibnamefont {Sierpowski}}, \bibinfo
  {author} {\bibfnamefont {S.}~\bibnamefont {Tashenov}}, \bibinfo {author}
  {\bibfnamefont {A.}~\bibnamefont {Warczak}},\ and\ \bibinfo {author}
  {\bibfnamefont {Y.}~\bibnamefont {Zou}},\ }\href
  {https://doi.org/10.1103/PhysRevLett.92.203004} {\bibfield  {journal}
  {\bibinfo  {journal} {Phys. Rev. Lett.}\ }\textbf {\bibinfo {volume} {92}},\
  \bibinfo {pages} {203004} (\bibinfo {year} {2004})}\BibitemShut {NoStop}%
\bibitem [{\citenamefont {Lestinsky}\ \emph {et~al.}(2022)\citenamefont
  {Lestinsky}, \citenamefont {Menz}, \citenamefont {Danared}, \citenamefont
  {Krantz}, \citenamefont {Lindroth}, \citenamefont {Andelkovic}, \citenamefont
  {Brandau}, \citenamefont {Br\"auning-Demian}, \citenamefont {Fedotova},
  \citenamefont {Geithner}, \citenamefont {Herfurth}, \citenamefont {Kalinin},
  \citenamefont {Kraus}, \citenamefont {Spillmann}, \citenamefont {Vorobyev},\
  and\ \citenamefont {St\"ohlker}}]{lestinsky_first_2022}%
  \BibitemOpen
  \bibfield  {author} {\bibinfo {author} {\bibfnamefont {M.}~\bibnamefont
  {Lestinsky}}, \bibinfo {author} {\bibfnamefont {E.~B.}\ \bibnamefont {Menz}},
  \bibinfo {author} {\bibfnamefont {H.}~\bibnamefont {Danared}}, \bibinfo
  {author} {\bibfnamefont {C.}~\bibnamefont {Krantz}}, \bibinfo {author}
  {\bibfnamefont {E.}~\bibnamefont {Lindroth}}, \bibinfo {author}
  {\bibfnamefont {Z.}~\bibnamefont {Andelkovic}}, \bibinfo {author}
  {\bibfnamefont {C.}~\bibnamefont {Brandau}}, \bibinfo {author} {\bibfnamefont
  {A.}~\bibnamefont {Br\"auning-Demian}}, \bibinfo {author} {\bibfnamefont
  {S.}~\bibnamefont {Fedotova}}, \bibinfo {author} {\bibfnamefont
  {W.}~\bibnamefont {Geithner}}, \bibinfo {author} {\bibfnamefont
  {F.}~\bibnamefont {Herfurth}}, \bibinfo {author} {\bibfnamefont
  {A.}~\bibnamefont {Kalinin}}, \bibinfo {author} {\bibfnamefont
  {I.}~\bibnamefont {Kraus}}, \bibinfo {author} {\bibfnamefont
  {U.}~\bibnamefont {Spillmann}}, \bibinfo {author} {\bibfnamefont
  {G.}~\bibnamefont {Vorobyev}},\ and\ \bibinfo {author} {\bibfnamefont
  {{\relax Th}.}~\bibnamefont {St\"ohlker}},\ }\href
  {https://doi.org/10.3390/atoms10040141} {\bibfield  {journal} {\bibinfo
  {journal} {Atoms}\ }\textbf {\bibinfo {volume} {10}},\ \bibinfo {pages} {141}
  (\bibinfo {year} {2022})}\BibitemShut {NoStop}%
\bibitem [{\citenamefont {Pies}\ \emph {et~al.}(2012)\citenamefont {Pies},
  \citenamefont {Sch\"afer}, \citenamefont {Heuser}, \citenamefont {Kempf},
  \citenamefont {Pabinger}, \citenamefont {Porst}, \citenamefont {Ranitsch},
  \citenamefont {Foerster}, \citenamefont {Hengstler}, \citenamefont
  {Kampk\"otter}, \citenamefont {Wolf}, \citenamefont {Gastaldo}, \citenamefont
  {Fleischmann},\ and\ \citenamefont {Enss}}]{Pies2012}%
  \BibitemOpen
  \bibfield  {author} {\bibinfo {author} {\bibfnamefont {C.}~\bibnamefont
  {Pies}}, \bibinfo {author} {\bibfnamefont {S.}~\bibnamefont {Sch\"afer}},
  \bibinfo {author} {\bibfnamefont {S.}~\bibnamefont {Heuser}}, \bibinfo
  {author} {\bibfnamefont {S.}~\bibnamefont {Kempf}}, \bibinfo {author}
  {\bibfnamefont {A.}~\bibnamefont {Pabinger}}, \bibinfo {author}
  {\bibfnamefont {J.~P.}\ \bibnamefont {Porst}}, \bibinfo {author}
  {\bibfnamefont {P.}~\bibnamefont {Ranitsch}}, \bibinfo {author}
  {\bibfnamefont {N.}~\bibnamefont {Foerster}}, \bibinfo {author}
  {\bibfnamefont {D.}~\bibnamefont {Hengstler}}, \bibinfo {author}
  {\bibfnamefont {A.}~\bibnamefont {Kampk\"otter}}, \bibinfo {author}
  {\bibfnamefont {T.}~\bibnamefont {Wolf}}, \bibinfo {author} {\bibfnamefont
  {L.}~\bibnamefont {Gastaldo}}, \bibinfo {author} {\bibfnamefont
  {A.}~\bibnamefont {Fleischmann}},\ and\ \bibinfo {author} {\bibfnamefont
  {C.}~\bibnamefont {Enss}},\ }\href
  {https://doi.org/10.1007/s10909-012-0557-z} {\bibfield  {journal} {\bibinfo
  {journal} {J. Low Temp. Phys.}\ }\textbf {\bibinfo {volume} {167}},\ \bibinfo
  {pages} {269} (\bibinfo {year} {2012})}\BibitemShut {NoStop}%
\bibitem [{\citenamefont {Hengstler}\ \emph {et~al.}(2015)\citenamefont
  {Hengstler}, \citenamefont {Keller}, \citenamefont {Sch\"otz}, \citenamefont
  {Geist}, \citenamefont {Krantz}, \citenamefont {Kempf}, \citenamefont
  {Gastaldo}, \citenamefont {Fleischmann}, \citenamefont {Gassner},
  \citenamefont {Weber}, \citenamefont {M\"artin}, \citenamefont {St\"ohlker},\
  and\ \citenamefont {Enss}}]{Hengstler2015}%
  \BibitemOpen
  \bibfield  {author} {\bibinfo {author} {\bibfnamefont {D.}~\bibnamefont
  {Hengstler}}, \bibinfo {author} {\bibfnamefont {M.}~\bibnamefont {Keller}},
  \bibinfo {author} {\bibfnamefont {C.}~\bibnamefont {Sch\"otz}}, \bibinfo
  {author} {\bibfnamefont {J.}~\bibnamefont {Geist}}, \bibinfo {author}
  {\bibfnamefont {M.}~\bibnamefont {Krantz}}, \bibinfo {author} {\bibfnamefont
  {S.}~\bibnamefont {Kempf}}, \bibinfo {author} {\bibfnamefont
  {L.}~\bibnamefont {Gastaldo}}, \bibinfo {author} {\bibfnamefont
  {A.}~\bibnamefont {Fleischmann}}, \bibinfo {author} {\bibfnamefont
  {T.}~\bibnamefont {Gassner}}, \bibinfo {author} {\bibfnamefont
  {G.}~\bibnamefont {Weber}}, \bibinfo {author} {\bibfnamefont
  {R.}~\bibnamefont {M\"artin}}, \bibinfo {author} {\bibfnamefont {{\relax
  Th}.}~\bibnamefont {St\"ohlker}},\ and\ \bibinfo {author} {\bibfnamefont
  {C.}~\bibnamefont {Enss}},\ }\href
  {https://doi.org/10.1088/0031-8949/2015/T166/014054} {\bibfield  {journal}
  {\bibinfo  {journal} {Phys. Scr.}\ }\textbf {\bibinfo {volume} {T166}},\
  \bibinfo {pages} {014054} (\bibinfo {year} {2015})}\BibitemShut {NoStop}%
\bibitem [{\citenamefont {Pf\"afflein}\ \emph
  {et~al.}(2022{\natexlab{a}})\citenamefont {Pf\"afflein}, \citenamefont
  {Allgeier}, \citenamefont {Bernitt}, \citenamefont {Fleischmann},
  \citenamefont {Friedrich}, \citenamefont {Hahn}, \citenamefont {Hengstler},
  \citenamefont {Herdrich}, \citenamefont {Kalinin}, \citenamefont {Kr\"oger},
  \citenamefont {Kuntz}, \citenamefont {Lestinsky}, \citenamefont {L\"oher},
  \citenamefont {Menz}, \citenamefont {Over}, \citenamefont {Spillmann},
  \citenamefont {Weber}, \citenamefont {Zhu}, \citenamefont {Enss},\ and\
  \citenamefont {St\"ohlker}}]{Pfaefflein2022}%
  \BibitemOpen
  \bibfield  {author} {\bibinfo {author} {\bibfnamefont {{\relax
  Ph}.}~\bibnamefont {Pf\"afflein}}, \bibinfo {author} {\bibfnamefont
  {S.}~\bibnamefont {Allgeier}}, \bibinfo {author} {\bibfnamefont
  {S.}~\bibnamefont {Bernitt}}, \bibinfo {author} {\bibfnamefont
  {A.}~\bibnamefont {Fleischmann}}, \bibinfo {author} {\bibfnamefont
  {M.}~\bibnamefont {Friedrich}}, \bibinfo {author} {\bibfnamefont
  {C.}~\bibnamefont {Hahn}}, \bibinfo {author} {\bibfnamefont {D.}~\bibnamefont
  {Hengstler}}, \bibinfo {author} {\bibfnamefont {M.~O.}\ \bibnamefont
  {Herdrich}}, \bibinfo {author} {\bibfnamefont {A.}~\bibnamefont {Kalinin}},
  \bibinfo {author} {\bibfnamefont {F.~M.}\ \bibnamefont {Kr\"oger}}, \bibinfo
  {author} {\bibfnamefont {P.}~\bibnamefont {Kuntz}}, \bibinfo {author}
  {\bibfnamefont {M.}~\bibnamefont {Lestinsky}}, \bibinfo {author}
  {\bibfnamefont {B.}~\bibnamefont {L\"oher}}, \bibinfo {author} {\bibfnamefont
  {E.~B.}\ \bibnamefont {Menz}}, \bibinfo {author} {\bibfnamefont
  {T.}~\bibnamefont {Over}}, \bibinfo {author} {\bibfnamefont {U.}~\bibnamefont
  {Spillmann}}, \bibinfo {author} {\bibfnamefont {G.}~\bibnamefont {Weber}},
  \bibinfo {author} {\bibfnamefont {B.}~\bibnamefont {Zhu}}, \bibinfo {author}
  {\bibfnamefont {C.}~\bibnamefont {Enss}},\ and\ \bibinfo {author}
  {\bibfnamefont {{\relax Th}.}~\bibnamefont {St\"ohlker}},\ }\href
  {https://doi.org/10.1088/1402-4896/ac93be} {\bibfield  {journal} {\bibinfo
  {journal} {Physica Scripta}\ }\textbf {\bibinfo {volume} {97}},\ \bibinfo
  {pages} {114005} (\bibinfo {year} {2022}{\natexlab{a}})}\BibitemShut
  {NoStop}%
\bibitem [{\citenamefont {Zhu}\ \emph {et~al.}(2022)\citenamefont {Zhu},
  \citenamefont {Gumberidze}, \citenamefont {Over}, \citenamefont {Weber},
  \citenamefont {Andelkovic}, \citenamefont {Br\"auning-Demian}, \citenamefont
  {Chen}, \citenamefont {Dmytriiev}, \citenamefont {Forstner}, \citenamefont
  {Hahn}, \citenamefont {Herfurth}, \citenamefont {Herdrich}, \citenamefont
  {Hillenbrand}, \citenamefont {Kalinin}, \citenamefont {Kr\"oger},
  \citenamefont {Lestinsky}, \citenamefont {Litvinov}, \citenamefont {Menz},
  \citenamefont {Middents}, \citenamefont {Morgenroth}, \citenamefont
  {Petridis}, \citenamefont {Pf\"afflein}, \citenamefont {Sanjari},
  \citenamefont {Sidhu}, \citenamefont {Spillmann}, \citenamefont {Schuch},
  \citenamefont {Schippers}, \citenamefont {Trotsenko}, \citenamefont {Varga},
  \citenamefont {Vorobyev},\ and\ \citenamefont {St\"ohlker}}]{Zhu2022}%
  \BibitemOpen
  \bibfield  {author} {\bibinfo {author} {\bibfnamefont {B.}~\bibnamefont
  {Zhu}}, \bibinfo {author} {\bibfnamefont {A.}~\bibnamefont {Gumberidze}},
  \bibinfo {author} {\bibfnamefont {T.}~\bibnamefont {Over}}, \bibinfo {author}
  {\bibfnamefont {G.}~\bibnamefont {Weber}}, \bibinfo {author} {\bibfnamefont
  {Z.}~\bibnamefont {Andelkovic}}, \bibinfo {author} {\bibfnamefont
  {A.}~\bibnamefont {Br\"auning-Demian}}, \bibinfo {author} {\bibfnamefont
  {R.~J.}\ \bibnamefont {Chen}}, \bibinfo {author} {\bibfnamefont
  {D.}~\bibnamefont {Dmytriiev}}, \bibinfo {author} {\bibfnamefont
  {O.}~\bibnamefont {Forstner}}, \bibinfo {author} {\bibfnamefont
  {C.}~\bibnamefont {Hahn}}, \bibinfo {author} {\bibfnamefont {F.}~\bibnamefont
  {Herfurth}}, \bibinfo {author} {\bibfnamefont {M.~O.}\ \bibnamefont
  {Herdrich}}, \bibinfo {author} {\bibfnamefont {P.-M.}\ \bibnamefont
  {Hillenbrand}}, \bibinfo {author} {\bibfnamefont {A.}~\bibnamefont
  {Kalinin}}, \bibinfo {author} {\bibfnamefont {F.~M.}\ \bibnamefont
  {Kr\"oger}}, \bibinfo {author} {\bibfnamefont {M.}~\bibnamefont {Lestinsky}},
  \bibinfo {author} {\bibfnamefont {Y.~A.}\ \bibnamefont {Litvinov}}, \bibinfo
  {author} {\bibfnamefont {E.~B.}\ \bibnamefont {Menz}}, \bibinfo {author}
  {\bibfnamefont {W.}~\bibnamefont {Middents}}, \bibinfo {author}
  {\bibfnamefont {T.}~\bibnamefont {Morgenroth}}, \bibinfo {author}
  {\bibfnamefont {N.}~\bibnamefont {Petridis}}, \bibinfo {author}
  {\bibfnamefont {{\relax Ph}.}~\bibnamefont {Pf\"afflein}}, \bibinfo {author}
  {\bibfnamefont {M.~S.}\ \bibnamefont {Sanjari}}, \bibinfo {author}
  {\bibfnamefont {R.~S.}\ \bibnamefont {Sidhu}}, \bibinfo {author}
  {\bibfnamefont {U.}~\bibnamefont {Spillmann}}, \bibinfo {author}
  {\bibfnamefont {R.}~\bibnamefont {Schuch}}, \bibinfo {author} {\bibfnamefont
  {S.}~\bibnamefont {Schippers}}, \bibinfo {author} {\bibfnamefont
  {S.}~\bibnamefont {Trotsenko}}, \bibinfo {author} {\bibfnamefont
  {L.}~\bibnamefont {Varga}}, \bibinfo {author} {\bibfnamefont
  {G.}~\bibnamefont {Vorobyev}},\ and\ \bibinfo {author} {\bibfnamefont
  {T.}~\bibnamefont {St\"ohlker}},\ }\href
  {https://doi.org/10.1103/PhysRevA.105.052804} {\bibfield  {journal} {\bibinfo
   {journal} {Phys. Rev. A}\ }\textbf {\bibinfo {volume} {105}},\ \bibinfo
  {pages} {052804} (\bibinfo {year} {2022})}\BibitemShut {NoStop}%
\bibitem [{\citenamefont {Kr\"oger}\ \emph {et~al.}(2023)\citenamefont
  {Kr\"oger}, \citenamefont {Weber}, \citenamefont {Allgeier}, \citenamefont
  {Andelkovic}, \citenamefont {Bernitt}, \citenamefont {Borovik}, \citenamefont
  {Duval}, \citenamefont {Fleischmann}, \citenamefont {Forstner}, \citenamefont
  {Friedrich}, \citenamefont {Glorius}, \citenamefont {Gumberidze},
  \citenamefont {Hahn}, \citenamefont {Herfurth}, \citenamefont {Hengstler},
  \citenamefont {Herdrich}, \citenamefont {Hillenbrand}, \citenamefont
  {Kalinin}, \citenamefont {Kiffer}, \citenamefont {Kubullek}, \citenamefont
  {Kuntz}, \citenamefont {Lestinsky}, \citenamefont {L\"oher}, \citenamefont
  {Menz}, \citenamefont {Over}, \citenamefont {Petridis}, \citenamefont
  {Pf\"afflein}, \citenamefont {Ringleb}, \citenamefont {Sidhu}, \citenamefont
  {Spillmann}, \citenamefont {Trotsenko}, \citenamefont {Warczak},
  \citenamefont {Zhu}, \citenamefont {Enss},\ and\ \citenamefont
  {St\"ohlker}}]{Kroeger23}%
  \BibitemOpen
  \bibfield  {author} {\bibinfo {author} {\bibfnamefont {F.~M.}\ \bibnamefont
  {Kr\"oger}}, \bibinfo {author} {\bibfnamefont {G.}~\bibnamefont {Weber}},
  \bibinfo {author} {\bibfnamefont {S.}~\bibnamefont {Allgeier}}, \bibinfo
  {author} {\bibfnamefont {Z.}~\bibnamefont {Andelkovic}}, \bibinfo {author}
  {\bibfnamefont {S.}~\bibnamefont {Bernitt}}, \bibinfo {author} {\bibfnamefont
  {A.}~\bibnamefont {Borovik}}, \bibinfo {author} {\bibfnamefont
  {L.}~\bibnamefont {Duval}}, \bibinfo {author} {\bibfnamefont
  {A.}~\bibnamefont {Fleischmann}}, \bibinfo {author} {\bibfnamefont
  {O.}~\bibnamefont {Forstner}}, \bibinfo {author} {\bibfnamefont
  {M.}~\bibnamefont {Friedrich}}, \bibinfo {author} {\bibfnamefont
  {J.}~\bibnamefont {Glorius}}, \bibinfo {author} {\bibfnamefont
  {A.}~\bibnamefont {Gumberidze}}, \bibinfo {author} {\bibfnamefont
  {C.}~\bibnamefont {Hahn}}, \bibinfo {author} {\bibfnamefont {F.}~\bibnamefont
  {Herfurth}}, \bibinfo {author} {\bibfnamefont {D.}~\bibnamefont {Hengstler}},
  \bibinfo {author} {\bibfnamefont {M.~O.}\ \bibnamefont {Herdrich}}, \bibinfo
  {author} {\bibfnamefont {P.-M.}\ \bibnamefont {Hillenbrand}}, \bibinfo
  {author} {\bibfnamefont {A.}~\bibnamefont {Kalinin}}, \bibinfo {author}
  {\bibfnamefont {M.}~\bibnamefont {Kiffer}}, \bibinfo {author} {\bibfnamefont
  {M.}~\bibnamefont {Kubullek}}, \bibinfo {author} {\bibfnamefont
  {P.}~\bibnamefont {Kuntz}}, \bibinfo {author} {\bibfnamefont
  {M.}~\bibnamefont {Lestinsky}}, \bibinfo {author} {\bibfnamefont
  {B.}~\bibnamefont {L\"oher}}, \bibinfo {author} {\bibfnamefont {E.~B.}\
  \bibnamefont {Menz}}, \bibinfo {author} {\bibfnamefont {T.}~\bibnamefont
  {Over}}, \bibinfo {author} {\bibfnamefont {N.}~\bibnamefont {Petridis}},
  \bibinfo {author} {\bibfnamefont {{\relax Ph}.}~\bibnamefont {Pf\"afflein}},
  \bibinfo {author} {\bibfnamefont {S.}~\bibnamefont {Ringleb}}, \bibinfo
  {author} {\bibfnamefont {R.~S.}\ \bibnamefont {Sidhu}}, \bibinfo {author}
  {\bibfnamefont {U.}~\bibnamefont {Spillmann}}, \bibinfo {author}
  {\bibfnamefont {S.}~\bibnamefont {Trotsenko}}, \bibinfo {author}
  {\bibfnamefont {A.}~\bibnamefont {Warczak}}, \bibinfo {author} {\bibfnamefont
  {B.}~\bibnamefont {Zhu}}, \bibinfo {author} {\bibfnamefont {C.}~\bibnamefont
  {Enss}},\ and\ \bibinfo {author} {\bibfnamefont {{\relax Th}.}~\bibnamefont
  {St\"ohlker}},\ }\bibfield  {journal} {\bibinfo  {journal} {Atoms}\ }\textbf
  {\bibinfo {volume} {11}},\ \href {https://doi.org/10.3390/atoms11020022}
  {10.3390/atoms11020022} (\bibinfo {year} {2023})\BibitemShut {NoStop}%
\bibitem [{\citenamefont {Gassner}\ \emph {et~al.}(2018)\citenamefont
  {Gassner}, \citenamefont {Trassinelli}, \citenamefont {Heß}, \citenamefont
  {Spillmann}, \citenamefont {Banaś}, \citenamefont {Blumenhagen},
  \citenamefont {Bosch}, \citenamefont {Brandau}, \citenamefont {Chen},
  \citenamefont {Chr}, \citenamefont {F\"orster}, \citenamefont {Grisenti},
  \citenamefont {Gumberidze}, \citenamefont {Hagmann}, \citenamefont
  {Hillenbrand}, \citenamefont {Indelicato}, \citenamefont {Jagodzinski},
  \citenamefont {K\"ampfer}, \citenamefont {Chr}, \citenamefont {Lestinsky},
  \citenamefont {Liesen}, \citenamefont {Yu}, \citenamefont {Loetzsch},
  \citenamefont {Manil}, \citenamefont {M\"artin}, \citenamefont {Nolden},
  \citenamefont {Petridis}, \citenamefont {Sanjari}, \citenamefont {Schulze},
  \citenamefont {Schwemlein}, \citenamefont {Simionovici}, \citenamefont
  {Steck}, \citenamefont {St\"ohlker}, \citenamefont {Szabo}, \citenamefont
  {Trotsenko}, \citenamefont {Uschmann}, \citenamefont {Weber}, \citenamefont
  {Wehrhan}, \citenamefont {Winckler}, \citenamefont {Winters}, \citenamefont
  {Winters}, \citenamefont {Ziegler},\ and\ \citenamefont
  {Beyer}}]{Gassner2018}%
  \BibitemOpen
  \bibfield  {author} {\bibinfo {author} {\bibfnamefont {T.}~\bibnamefont
  {Gassner}}, \bibinfo {author} {\bibfnamefont {M.}~\bibnamefont
  {Trassinelli}}, \bibinfo {author} {\bibfnamefont {R.}~\bibnamefont {Heß}},
  \bibinfo {author} {\bibfnamefont {U.}~\bibnamefont {Spillmann}}, \bibinfo
  {author} {\bibfnamefont {D.}~\bibnamefont {Banaś}}, \bibinfo {author}
  {\bibfnamefont {K.~H.}\ \bibnamefont {Blumenhagen}}, \bibinfo {author}
  {\bibfnamefont {F.}~\bibnamefont {Bosch}}, \bibinfo {author} {\bibfnamefont
  {C.}~\bibnamefont {Brandau}}, \bibinfo {author} {\bibfnamefont
  {W.}~\bibnamefont {Chen}}, \bibinfo {author} {\bibfnamefont {D.}~\bibnamefont
  {Chr}}, \bibinfo {author} {\bibfnamefont {E.}~\bibnamefont {F\"orster}},
  \bibinfo {author} {\bibfnamefont {R.~E.}\ \bibnamefont {Grisenti}}, \bibinfo
  {author} {\bibfnamefont {A.}~\bibnamefont {Gumberidze}}, \bibinfo {author}
  {\bibfnamefont {S.}~\bibnamefont {Hagmann}}, \bibinfo {author} {\bibfnamefont
  {P.~M.}\ \bibnamefont {Hillenbrand}}, \bibinfo {author} {\bibfnamefont
  {P.}~\bibnamefont {Indelicato}}, \bibinfo {author} {\bibfnamefont
  {P.}~\bibnamefont {Jagodzinski}}, \bibinfo {author} {\bibfnamefont
  {T.}~\bibnamefont {K\"ampfer}}, \bibinfo {author} {\bibfnamefont
  {K.}~\bibnamefont {Chr}}, \bibinfo {author} {\bibfnamefont {M.}~\bibnamefont
  {Lestinsky}}, \bibinfo {author} {\bibfnamefont {D.}~\bibnamefont {Liesen}},
  \bibinfo {author} {\bibfnamefont {A.~L.}\ \bibnamefont {Yu}}, \bibinfo
  {author} {\bibfnamefont {R.}~\bibnamefont {Loetzsch}}, \bibinfo {author}
  {\bibfnamefont {B.}~\bibnamefont {Manil}}, \bibinfo {author} {\bibfnamefont
  {R.}~\bibnamefont {M\"artin}}, \bibinfo {author} {\bibfnamefont
  {F.}~\bibnamefont {Nolden}}, \bibinfo {author} {\bibfnamefont
  {N.}~\bibnamefont {Petridis}}, \bibinfo {author} {\bibfnamefont {M.~S.}\
  \bibnamefont {Sanjari}}, \bibinfo {author} {\bibfnamefont {K.~S.}\
  \bibnamefont {Schulze}}, \bibinfo {author} {\bibfnamefont {M.}~\bibnamefont
  {Schwemlein}}, \bibinfo {author} {\bibfnamefont {A.}~\bibnamefont
  {Simionovici}}, \bibinfo {author} {\bibfnamefont {M.}~\bibnamefont {Steck}},
  \bibinfo {author} {\bibfnamefont {{\relax Th}.}~\bibnamefont {St\"ohlker}},
  \bibinfo {author} {\bibfnamefont {C.~I.}\ \bibnamefont {Szabo}}, \bibinfo
  {author} {\bibfnamefont {S.}~\bibnamefont {Trotsenko}}, \bibinfo {author}
  {\bibfnamefont {I.}~\bibnamefont {Uschmann}}, \bibinfo {author}
  {\bibfnamefont {G.}~\bibnamefont {Weber}}, \bibinfo {author} {\bibfnamefont
  {O.}~\bibnamefont {Wehrhan}}, \bibinfo {author} {\bibfnamefont
  {N.}~\bibnamefont {Winckler}}, \bibinfo {author} {\bibfnamefont {D.~F.~A.}\
  \bibnamefont {Winters}}, \bibinfo {author} {\bibfnamefont {N.}~\bibnamefont
  {Winters}}, \bibinfo {author} {\bibfnamefont {E.}~\bibnamefont {Ziegler}},\
  and\ \bibinfo {author} {\bibfnamefont {H.~F.}\ \bibnamefont {Beyer}},\ }\href
  {https://doi.org/10.1088/1367-2630/aad01d} {\bibfield  {journal} {\bibinfo
  {journal} {New J. Phys.}\ }\textbf {\bibinfo {volume} {20}},\ \bibinfo
  {pages} {073033} (\bibinfo {year} {2018})}\BibitemShut {NoStop}%
\bibitem [{\citenamefont {Kempf}\ \emph {et~al.}(2018)\citenamefont {Kempf},
  \citenamefont {Fleischmann}, \citenamefont {Gastaldo},\ and\ \citenamefont
  {Enss}}]{Kempf2018}%
  \BibitemOpen
  \bibfield  {author} {\bibinfo {author} {\bibfnamefont {S.}~\bibnamefont
  {Kempf}}, \bibinfo {author} {\bibfnamefont {A.}~\bibnamefont {Fleischmann}},
  \bibinfo {author} {\bibfnamefont {L.}~\bibnamefont {Gastaldo}},\ and\
  \bibinfo {author} {\bibfnamefont {C.}~\bibnamefont {Enss}},\ }\href
  {https://doi.org/10.1007/s10909-018-1891-6} {\bibfield  {journal} {\bibinfo
  {journal} {J. Low Temp. Phys.}\ }\textbf {\bibinfo {volume} {193}},\ \bibinfo
  {pages} {365} (\bibinfo {year} {2018})}\BibitemShut {NoStop}%
\bibitem [{\citenamefont {Pf\"afflein}\ \emph
  {et~al.}(2022{\natexlab{b}})\citenamefont {Pf\"afflein}, \citenamefont
  {Weber}, \citenamefont {Allgeier}, \citenamefont {Bernitt}, \citenamefont
  {Fleischmann}, \citenamefont {Friedrich}, \citenamefont {Hahn}, \citenamefont
  {Hengstler}, \citenamefont {Herdrich}, \citenamefont {Kalinin}, \citenamefont
  {Kr\"oger}, \citenamefont {Kuntz}, \citenamefont {Lestinsky}, \citenamefont
  {L\"oher}, \citenamefont {Menz}, \citenamefont {Spillmann}, \citenamefont
  {Zhu}, \citenamefont {Enss},\ and\ \citenamefont
  {St\"ohlker}}]{Pfaefflein2022_coinc}%
  \BibitemOpen
  \bibfield  {author} {\bibinfo {author} {\bibfnamefont {{\relax
  Ph}.}~\bibnamefont {Pf\"afflein}}, \bibinfo {author} {\bibfnamefont
  {G.}~\bibnamefont {Weber}}, \bibinfo {author} {\bibfnamefont
  {S.}~\bibnamefont {Allgeier}}, \bibinfo {author} {\bibfnamefont
  {S.}~\bibnamefont {Bernitt}}, \bibinfo {author} {\bibfnamefont
  {A.}~\bibnamefont {Fleischmann}}, \bibinfo {author} {\bibfnamefont
  {M.}~\bibnamefont {Friedrich}}, \bibinfo {author} {\bibfnamefont
  {C.}~\bibnamefont {Hahn}}, \bibinfo {author} {\bibfnamefont {D.}~\bibnamefont
  {Hengstler}}, \bibinfo {author} {\bibfnamefont {M.~O.}\ \bibnamefont
  {Herdrich}}, \bibinfo {author} {\bibfnamefont {A.}~\bibnamefont {Kalinin}},
  \bibinfo {author} {\bibfnamefont {F.~M.}\ \bibnamefont {Kr\"oger}}, \bibinfo
  {author} {\bibfnamefont {P.}~\bibnamefont {Kuntz}}, \bibinfo {author}
  {\bibfnamefont {M.}~\bibnamefont {Lestinsky}}, \bibinfo {author}
  {\bibfnamefont {B.}~\bibnamefont {L\"oher}}, \bibinfo {author} {\bibfnamefont
  {E.~B.}\ \bibnamefont {Menz}}, \bibinfo {author} {\bibfnamefont
  {U.}~\bibnamefont {Spillmann}}, \bibinfo {author} {\bibfnamefont
  {B.}~\bibnamefont {Zhu}}, \bibinfo {author} {\bibfnamefont {C.}~\bibnamefont
  {Enss}},\ and\ \bibinfo {author} {\bibfnamefont {{\relax Th}.}~\bibnamefont
  {St\"ohlker}},\ }\href {https://doi.org/10.3390/atoms11010005} {\bibfield
  {journal} {\bibinfo  {journal} {Atoms}\ }\textbf {\bibinfo {volume} {11}},\
  \bibinfo {pages} {5} (\bibinfo {year} {2022}{\natexlab{b}})}\BibitemShut
  {NoStop}%
\bibitem [{\citenamefont {Bates}\ \emph {et~al.}(2016)\citenamefont {Bates},
  \citenamefont {Pies}, \citenamefont {Kempf}, \citenamefont {Hengstler},
  \citenamefont {Fleischmann}, \citenamefont {Gastaldo}, \citenamefont {Enss},\
  and\ \citenamefont {Friedrich}}]{Bates2016}%
  \BibitemOpen
  \bibfield  {author} {\bibinfo {author} {\bibfnamefont {C.~R.}\ \bibnamefont
  {Bates}}, \bibinfo {author} {\bibfnamefont {C.}~\bibnamefont {Pies}},
  \bibinfo {author} {\bibfnamefont {S.}~\bibnamefont {Kempf}}, \bibinfo
  {author} {\bibfnamefont {D.}~\bibnamefont {Hengstler}}, \bibinfo {author}
  {\bibfnamefont {A.}~\bibnamefont {Fleischmann}}, \bibinfo {author}
  {\bibfnamefont {L.}~\bibnamefont {Gastaldo}}, \bibinfo {author}
  {\bibfnamefont {C.}~\bibnamefont {Enss}},\ and\ \bibinfo {author}
  {\bibfnamefont {S.}~\bibnamefont {Friedrich}},\ }\href
  {https://doi.org/10.1063/1.4958699} {\bibfield  {journal} {\bibinfo
  {journal} {Appl. Phys. Lett.}\ }\textbf {\bibinfo {volume} {109}},\ \bibinfo
  {pages} {023513} (\bibinfo {year} {2016})}\BibitemShut {NoStop}%
\bibitem [{\citenamefont {Helmer}\ and\ \citenamefont {van~der
  Leun}(2000)}]{Helmer2000}%
  \BibitemOpen
  \bibfield  {author} {\bibinfo {author} {\bibfnamefont {R.}~\bibnamefont
  {Helmer}}\ and\ \bibinfo {author} {\bibfnamefont {C.}~\bibnamefont {van~der
  Leun}},\ }\href {https://doi.org/10.1016/S0168-9002(00)00252-7} {\bibfield
  {journal} {\bibinfo  {journal} {Nucl. Instrum. Meth. A}\ }\textbf {\bibinfo
  {volume} {450}},\ \bibinfo {pages} {35} (\bibinfo {year} {2000})}\BibitemShut
  {NoStop}%
\bibitem [{\citenamefont {James}\ and\ \citenamefont
  {Roos}(1975)}]{James:1975dr}%
  \BibitemOpen
  \bibfield  {author} {\bibinfo {author} {\bibfnamefont {F.}~\bibnamefont
  {James}}\ and\ \bibinfo {author} {\bibfnamefont {M.}~\bibnamefont {Roos}},\
  }\href {https://doi.org/10.1016/0010-4655(75)90039-9} {\bibfield  {journal}
  {\bibinfo  {journal} {Comput. Phys. Commun.}\ }\textbf {\bibinfo {volume}
  {10}},\ \bibinfo {pages} {343} (\bibinfo {year} {1975})}\BibitemShut
  {NoStop}%
\bibitem [{\citenamefont {Dembinski}\ and\ \citenamefont
  {et~al.}(2020)}]{iminuit}%
  \BibitemOpen
  \bibfield  {author} {\bibinfo {author} {\bibfnamefont {H.}~\bibnamefont
  {Dembinski}}\ and\ \bibinfo {author} {\bibfnamefont {P.~O.}\ \bibnamefont
  {et~al.}}\ }\href {https://doi.org/10.5281/zenodo.3949207}
  {10.5281/zenodo.3949207} (\bibinfo {year} {2020})\BibitemShut {NoStop}%
\bibitem [{\citenamefont {Thorn}\ \emph {et~al.}(2008)\citenamefont {Thorn},
  \citenamefont {Brown}, \citenamefont {Clementson}, \citenamefont {Chen},
  \citenamefont {Chen}, \citenamefont {Beiersdorfer}, \citenamefont {Boyce},
  \citenamefont {Kilbourne}, \citenamefont {Porter},\ and\ \citenamefont
  {Kelley}}]{thorn_high-resolution_2008}%
  \BibitemOpen
  \bibfield  {author} {\bibinfo {author} {\bibfnamefont {D.~B.}\ \bibnamefont
  {Thorn}}, \bibinfo {author} {\bibfnamefont {G.~V.}\ \bibnamefont {Brown}},
  \bibinfo {author} {\bibfnamefont {J.~H.}\ \bibnamefont {Clementson}},
  \bibinfo {author} {\bibfnamefont {H.}~\bibnamefont {Chen}}, \bibinfo {author}
  {\bibfnamefont {M.}~\bibnamefont {Chen}}, \bibinfo {author} {\bibfnamefont
  {P.}~\bibnamefont {Beiersdorfer}}, \bibinfo {author} {\bibfnamefont {K.~R.}\
  \bibnamefont {Boyce}}, \bibinfo {author} {\bibfnamefont {C.~A.}\ \bibnamefont
  {Kilbourne}}, \bibinfo {author} {\bibfnamefont {F.~S.}\ \bibnamefont
  {Porter}},\ and\ \bibinfo {author} {\bibfnamefont {R.~L.}\ \bibnamefont
  {Kelley}},\ }\href {https://doi.org/10.1139/p07-134} {\bibfield  {journal}
  {\bibinfo  {journal} {Canadian Journal of Physics}\ }\textbf {\bibinfo
  {volume} {86}},\ \bibinfo {pages} {241} (\bibinfo {year} {2008})}\BibitemShut
  {NoStop}%
\bibitem [{\citenamefont {Lupton}\ \emph {et~al.}(1994)\citenamefont {Lupton},
  \citenamefont {Dietrich}, \citenamefont {Hailey}, \citenamefont {Stewart},\
  and\ \citenamefont {Ziock}}]{Lupton94}%
  \BibitemOpen
  \bibfield  {author} {\bibinfo {author} {\bibfnamefont {J.~H.}\ \bibnamefont
  {Lupton}}, \bibinfo {author} {\bibfnamefont {D.~D.}\ \bibnamefont
  {Dietrich}}, \bibinfo {author} {\bibfnamefont {C.~J.}\ \bibnamefont
  {Hailey}}, \bibinfo {author} {\bibfnamefont {R.~E.}\ \bibnamefont
  {Stewart}},\ and\ \bibinfo {author} {\bibfnamefont {K.~P.}\ \bibnamefont
  {Ziock}},\ }\href {https://doi.org/10.1103/PhysRevA.50.2150} {\bibfield
  {journal} {\bibinfo  {journal} {Phys. Rev. A}\ }\textbf {\bibinfo {volume}
  {50}},\ \bibinfo {pages} {2150} (\bibinfo {year} {1994})}\BibitemShut
  {NoStop}%
\bibitem [{\citenamefont {Gumberidze}\ \emph {et~al.}(2005)\citenamefont
  {Gumberidze}, \citenamefont {St\"ohlker}, \citenamefont {Banas},
  \citenamefont {Beckert}, \citenamefont {Beller}, \citenamefont {Beyer},
  \citenamefont {Bosch}, \citenamefont {Hagmann}, \citenamefont {Kozhuharov},
  \citenamefont {Liesen}, \citenamefont {Nolden}, \citenamefont {Ma},
  \citenamefont {Mokler}, \citenamefont {Steck}, \citenamefont {Sierpowski},\
  and\ \citenamefont {Tashenov}}]{Gumberidze2005}%
  \BibitemOpen
  \bibfield  {author} {\bibinfo {author} {\bibfnamefont {A.}~\bibnamefont
  {Gumberidze}}, \bibinfo {author} {\bibfnamefont {{\relax Th}.}~\bibnamefont
  {St\"ohlker}}, \bibinfo {author} {\bibfnamefont {D.}~\bibnamefont {Banas}},
  \bibinfo {author} {\bibfnamefont {K.}~\bibnamefont {Beckert}}, \bibinfo
  {author} {\bibfnamefont {P.}~\bibnamefont {Beller}}, \bibinfo {author}
  {\bibfnamefont {H.~F.}\ \bibnamefont {Beyer}}, \bibinfo {author}
  {\bibfnamefont {F.}~\bibnamefont {Bosch}}, \bibinfo {author} {\bibfnamefont
  {S.}~\bibnamefont {Hagmann}}, \bibinfo {author} {\bibfnamefont
  {C.}~\bibnamefont {Kozhuharov}}, \bibinfo {author} {\bibfnamefont
  {D.}~\bibnamefont {Liesen}}, \bibinfo {author} {\bibfnamefont
  {F.}~\bibnamefont {Nolden}}, \bibinfo {author} {\bibfnamefont
  {X.}~\bibnamefont {Ma}}, \bibinfo {author} {\bibfnamefont {P.~H.}\
  \bibnamefont {Mokler}}, \bibinfo {author} {\bibfnamefont {M.}~\bibnamefont
  {Steck}}, \bibinfo {author} {\bibfnamefont {D.}~\bibnamefont {Sierpowski}},\
  and\ \bibinfo {author} {\bibfnamefont {S.}~\bibnamefont {Tashenov}},\ }\href
  {https://doi.org/10.1103/PhysRevLett.94.223001} {\bibfield  {journal}
  {\bibinfo  {journal} {Phys. Rev. Lett.}\ }\textbf {\bibinfo {volume} {94}},\
  \bibinfo {pages} {223001} (\bibinfo {year} {2005})}\BibitemShut {NoStop}%
\end{thebibliography}%

\end{document}